\documentclass[10pt,journal,compsoc]{IEEEtran}
\ifCLASSOPTIONcompsoc
  \usepackage[nocompress]{cite}
\else
  \usepackage{cite}
\fi
\ifCLASSINFOpdf
\else
\fi
\usepackage{graphicx}
\usepackage{multirow}
\usepackage{times}
\usepackage{amsmath,amsthm}
\usepackage{amssymb}
\usepackage{amsfonts}
\usepackage{algorithm}
\usepackage{mathrsfs}
\usepackage{booktabs} % For formal tables
\usepackage{color}
\usepackage{hyperref} 
\usepackage{url}
\usepackage{algpseudocode}
\usepackage{pifont}
\usepackage{bm}
\usepackage{comment}
\usepackage{wasysym}
\usepackage{bbding-zy}

\def\1{{\bf{1}}}
\def\0{{\bf{0}}}

\def\b{{\bf b}}

\def\d{{\bf d}}
\def\e{{\bf e}}

\def\h{{\bf h}}

\def\j{{\bf j}} 
\def\k{{\bf k}}

\def\q{{\bf q}}
\def\r{{\bf r}}
\def\s{{\bf s}}

\def\u{{\bf u}}

\def\z{{\bf z}}

\def\D{{\bf D}}

\def\H{{\bf H}}

\def\W{{\bf W}}

\def\Acal{{\mathcal{A}}}
\def\Bcal{{\mathcal{B}}}

\def\Ecal{{\mathcal{E}}}

\def\Gcal{{\mathcal{G}}}

\def\Jcal{{\mathcal{J}}}

\def\Lcal{{\mathcal{L}}}

\def\Ncal{{\mathcal{N}}}

\def\Rcal{{\mathcal{R}}}

\def\Tcal{{\mathcal{T}}}

\def\Vcal{{\mathcal{V}}}

\def\Ycal{{\mathcal{Y}}}

\def\Rbb{{\mathbb R}}

\newtheorem{definition}{\textbf{Definition}}
\newtheorem{problem}{\textbf{Problem}}

\begin{document}
% \title{Bare Advanced Demo of IEEEtran.cls for\\ IEEE Computer Society Journals}
% \title{ConnectE-MRGAT: Exploring Neighborhood Information for Knowledge Graph Entity Typing}
% \title{Representation Learning on Bi-typed Heterogeneous Graph via Dual Hierarchical Attention Networks}
\title{Combining Intra-Risk and Contagion Risk for Enterprise Bankruptcy Prediction Using Graph Neural Networks}

\author{Yu~Zhao,
        Shaopeng~Wei,
        Yu~Guo,
        Qing~Yang,
        Xingyan~Chen,
        Qing~Li,~\IEEEmembership{Member,~IEEE},
        Fuzhen~Zhuang,~\IEEEmembership{Member,~IEEE},
        Ji~Liu~\IEEEmembership{Member,~IEEE},
        % Ce Zhang,
        Gang~Kou~\IEEEmembership{Member,~IEEE}
        % <-this % stops a space
\IEEEcompsocitemizethanks{
\IEEEcompsocthanksitem Y. Zhao, Y. Guo, Q. Yang, X. Chen and Q. Li are with Financial Intelligence and Financial Engineering Key Laboratory of Sichuan Province, Department of Artificial Intelligence, Southwestern University of Finance and Economics, China.\protect\\
E-mail: zhaoyu@swufe.edu.cn

\IEEEcompsocthanksitem S. Wei and G. Kou are with School of Business Administration, Faculty of Business Administration, Southwestern University of Finance and Economics, Chengdu, 611130, China.
\protect\\
E-mail: kougang@swufe.edu.cn 

\IEEEcompsocthanksitem F. Zhuang is with Institute of Artificial Intelligence, Beihang University, Beijing 100191, China; and with 
SKLSDE, School of Computer Science, Beihang University, Beijing 100191, China \protect\\
E-mail:zhuangfuzhen@buaa.edu.cn 

\IEEEcompsocthanksitem J. Liu is with Kuaishou Technology, USA.\protect\\
E-mail:ji.liu.uwisc@gmail.com
% \IEEEcompsocthanksitem C. Zhang is with ETH Zürich, Switzerland. \protect\\
% E-mail:ce.zhang@inf.ethz.ch
\IEEEcompsocthanksitem G. Kou is the corresponding author.

% \IEEEcompsocthanksitem X. Wang and J. Guo are with the School of Artifical Intelligence, Beijing University of Posts and Telecommunications, China.
% \IEEEcompsocthanksitem Jun Guo is with .
}% <-this % stops a space
% \thanks{Manuscript received April 19, 2005; revised August 26, 2015.}
}

\IEEEtitleabstractindextext{%
\begin{abstract}
  % \edit{Bankruptcy prediction?}

% \textbf{Enterprise Bankruptcy prediction} aims at predicting listed companies' stock future price trend, which is a challenging task due to the volatile nature of financial markets. 

Predicting the bankruptcy risk of small and medium-sized enterprises (SMEs) is an important step for financial institutions when making decisions about loans.
% the loan decision and identify region economics's early warning. 
% \edit{How previous methods solve it? What's the challenge?} 
Existing studies in both finance and AI research fields, however, tend to only consider either the intra-risk or contagion risk of enterprises, ignoring their interactions and combinatorial effects. 
%to sufficiently model enterprise risk as a combination result of multi-source intra-risk, and complex risk momentum spillover effect. 
% \textbf{H\superscript{2}GNN}
% \edit{How we do? What's the main difference? Why our method works?} 
This study for the first time considers both types of risk and their joint effects in bankruptcy prediction.
%In this paper, we propose an approach to predict enterprise bankruptcy risk by combining enterprise intra-risk and  contagion risk. 
% which is trained by jointly utilizing (i) , and (ii) 
% \textbf{In this paper, we propose a novel risk fusing network to model enterprise risk as the combination of inner risk and transmission risk. }
Specifically, we first propose an enterprise intra-risk encoder based on statistically significant enterprise risk indicators taken from its basic business information as well as litigation information for its intra-risk learning. 
% to mining enterprise inner risk, which consists of SMEs basic business information and lawsuit information. 
Then, we propose an enterprise contagion risk encoder based on enterprise relation information from an enterprise knowledge graph for its contagion risk embedding. In particular, the contagion risk encoder includes both the newly proposed Hyper-Graph Neural Networks (Hyper-GNNs) and Heterogeneous Graph Neural Networks (Heter-GNNs), which can model contagion risk in two different aspects, i.e. common risk factors based on hyperedges and direct diffusion risk from neighbors, respectively. 
% hyperedge and multiplex heterogeneous relations among enterprise connected graph, respectively.
% Secondly, we conduct a hierarchical transformer encoder to mining transmission risk in enterprise knowledge graphs. 
% Specifically, we conduct weighted averaging and attention mechanism for weighted and unweighted multi-relations respectively in entities level information aggregation. 
% Then we perform transformer based aggregation mechanism to merge risk route level information. 
% Thirdly, we utilize a hierarchical hypergraph encoder to model high-order relations among enterprises.
Using these two types of encoders, we design a unified framework to simultaneously capture intra-risk and contagion risk for bankruptcy prediction. 
To evaluate the model, we collect real-world multi-sources data on SMEs and build a novel benchmark dataset called SMEsD. We provide open access to the dataset, which is expected to further promote research on financial risk analysis. 
Experiments on SMEsD against twelve state-of-the-art baselines demonstrate the effectiveness of the proposed model for bankruptcy prediction.
% Experimental results demonstrate that the proposed risk fusing network outperforms state-of-the-art methods.
\end{abstract}

\begin{IEEEkeywords}
% Computer Society, IEEE, IEEEtran, journal, \LaTeX, paper, template.
Enterprise Bankruptcy Prediction, Intra-Risk, Contagion Risk, Hyper-GNNs, Heterogeneous GNNs
\end{IEEEkeywords}}

\maketitle

% is not selected <OR> if conference mode is selected - because compsoc
% conference papers position the abstract like regular (non-compsoc)
\IEEEdisplaynontitleabstractindextext

% \fi
\IEEEpeerreviewmaketitle

\ifCLASSOPTIONcompsoc
\IEEEraisesectionheading{\section{Introduction}\label{sec:introduction}}
\else

\section{Introduction}
\label{sec:introduction}
\fi
\IEEEPARstart{S}{mall} and medium-sized enterprises (SMEs) contribute up to $40\%$ of gross domestic product (GDP) in emerging economies and provide more than $50\%$ of employment worldwide\footnote{\url{https://www.worldbank.org/en/topic/smefinance}}. Predicting the financial risk of SMEs is of great importance for both government policymakers and financial institutions \cite{kongolo2010job,moro2013loan}. 
% \edit{How previous methods solve it? However, What's the challenge? point 1 and point 2;}
Previous studies of enterprise risk in both finance and AI research fields typically either examine enterprises’ internal financial aspects to detect intra-risk (i.e., risks resulting from enterprises' operations), or they analyze risk diffusion based on simulations
 \cite{odom1990neural,min2009binary,li2020maec,hoberg2015redefining,liu2021finbert,basole2014supply,rahmandad2008heterogeneity} to mine contagion risk (i.e., risk from external stakeholders), including upstream and downstream companies and related persons.
For simplicity, however, most studies only consider either \textbf{intra-risk} or \textbf{contagion risk} individually, ignoring their interactions and combinatorial effects.
% However, SMEs financial risk prediction in real market scenarios remains challenging because of the multi-source heterogeneity characteristics of the intra-risk data and the multiplex contagion risk relations of enterprises \cite{pan2021heterogeneous}. 
% Moreover, different from list firms financial analysis which can collect enough opening data from financial market, SMEs risk studies always face data deficiency \cite{yang2020financial}. 
% It is a non-trivial and challenging task to build up a uniform framework for enterprise bankruptcy prediction, taking advantage of both enterprise intra-risk and contagion risk due to the multi-source heterogeneity characteristics of the intra-risk data and the multiplex contagion risk relations of enterprises \cite{pan2021heterogeneous}.
Given the heterogeneous multi-source characteristics of intra-risk data and the complexities of contagion risk relations among enterprises, building a framework for enterprise bankruptcy prediction that considers both intra-risk and contagion risk is a non-trivial and challenging task.
% the heterogeneity of the SMEs multi-source data.

\begin{figure}[t]
    \centering
    \includegraphics[width=0.4\textwidth]{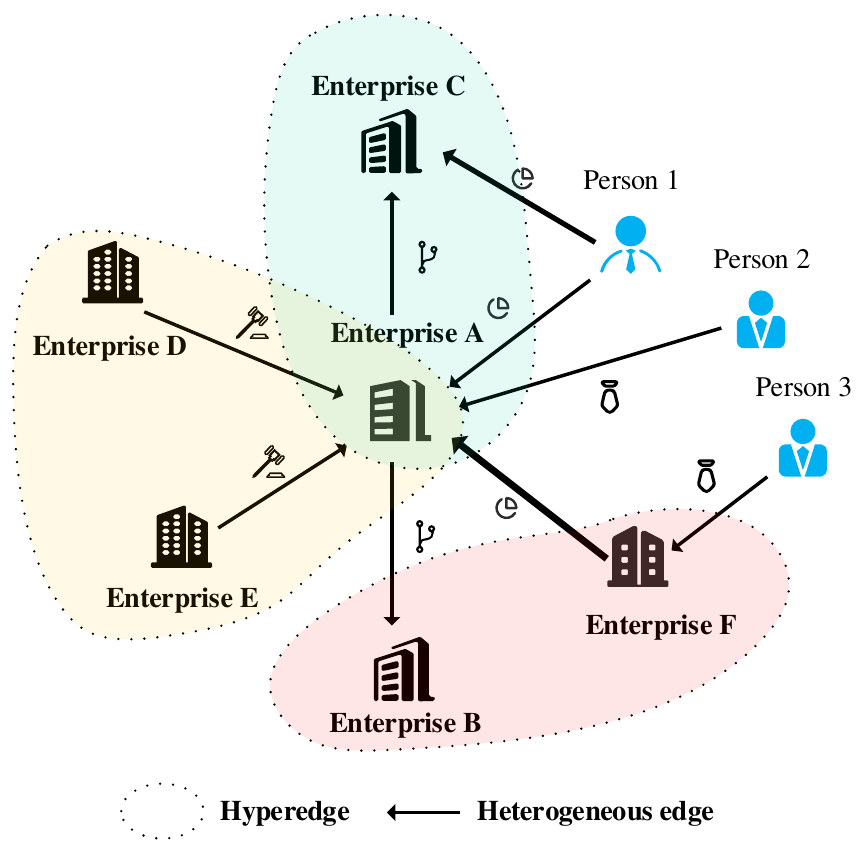}
    \caption{A toy example of enterprise knowledge graph which is extracted from the newly constructed dataset SMEsD.}
    \label{fig:example}
\end{figure}

To meet this challenge, we propose a novel enterprise bankruptcy prediction method that combines intra-risk with contagion risk. 
First, we propose an intra-risk encoder that leverages rich features based on enterprises' basic business information and litigation information to mine intra-risk. After statistically analyzing the correlations between enterprises' basic intelligence (including basic business attributes and litigation information ) and their bankruptcy risk (see Table \ref{tab:bankruptcy-analysis}), we successfully select 12 statistically significant indices for intra-risk encoder learning (see Section \ref{section-correlation-analysis} for details about the analysis of statistical significance). 
\begin{table*}[htb]
\caption{The statistical significance analysis on the correlation between the enterprises' basic intelligence (including the enterprise basic attributes, the enterprise litigation information here) and their bankruptcy risk. The symbols ***, ** and * denote the statistical result is significant in 99\%, 95\% and 90\% level, respectively.
% ** denotes the statistic result is significant in 95\% level and * denotes the statistic result is significant in 90\% level.
}
\label{tab:bankruptcy-analysis}
\newcommand{\tabincell}[2]{\begin{tabular}{@{}#1@{}}#2\end{tabular}}
\resizebox{\linewidth}{!}{
\begin{tabular}{l|l|cc|c|c|c}
\toprule
\multirow{2}{*}{ \textbf{Enterprises intelligence}  }  & \multirow{2}{*}{\textbf{Significant indices}} & \multicolumn{2}{c|}{\textbf{Correlation analysis} }  & \multicolumn{3}{c}{\textbf{Independent Samples t Test}} \\
\cline{3-7}
 &&\tabincell{c}{  \textbf{Coefficient} }& \textbf{Polarity}&\tabincell{c}{\textbf{Average number of }\\ \textbf{surviving enterprises}} & \tabincell{c}{\textbf{Average number of }\\ \textbf{bankrupted enterprises}}& \tabincell{c}{\textbf{Significance value of}\\ \textbf{average difference}} \\
\midrule
\midrule
\multirow{3}{*}{Enterprise Attributes} & Established time  & -.058*** & Negative & 156  & 148 & .000***            \\
                    & Registered capital & -.187*** & Negative & 16874 & 910  & .016**            \\
 & Paid-in capital    & -.159*** & Negative & 16264 & 873  & .020**            \\ 
\midrule
\midrule
\multirow{2}{*}{Lawsuit Cause}    & Loan contract dispute         & .122*** & Positive  & 1.80 & 2.23 & .032**            \\
& Sales contract dispute        & .077*** & Positive & .55   & .80  & .000***           \\ 
\midrule
\multirow{3}{*}{Court Level of Lawsuit}     & Grassroots people's court  & .086*** & Positive  &2.79  & 3.43 & .019**            \\
 & Intermediate people's court         & -.029** & Negative & .59   & .43  & .012**    \\      
 & Higher people's court         & -.070*** & Negative & .05   & .01  & .000***           \\ 
\midrule
\multirow{2}{*}{Verdict}    & Plaintiff winner       &-.076*** & Negative & .88   & .24  & .000***           \\
 & Defendant loser        & .124***  & Positive & 1.87  & 3.12 & .000***           \\ 
\midrule
\multirow{2}{*}{Duration Of Action}      & Less than two years           & .079*** & Positive & 3.23  & 3.82 & .059*             \\
  & More than two years         &-.086*** & Negative & .19   & .06  & .000***           \\ 
\bottomrule
\end{tabular}}
\end{table*}
Second, we propose an enterprise contagion risk encoder based on enterprise relational information from an enterprise knowledge graph (EKG) to embed contagion risk (also known as "risk momentum spillover effect" \cite{jegadeesh1993returns}).
Figure \ref{fig:example} shows a toy example of an EKG, from which we can find that enterprises have two kinds of relations: hyperedges and pair-wise heterogeneous relations (see Section \ref{section-risk-momentum-contagion} for details). 
Hence, we equip the contagion risk encoder with two submodels—hypergraph neural networks (Hyper-GNNs) and heterogeneous graph neural networks (Heter-GNNs)—to model risk diffusion in the EKG. Specifically, Hyper-GNNs aims to mine hyperedges in the EKG, such as the same industry and the same area, which is beneficial for enterprise risk prediction. 
During COVID-19, for example, most mask and vaccine manufacturers in the medical industry experienced a boom while the catering industry faced a significant bankruptcy risk. Heter-GNNs, meanwhile, can capture the direct contagion risk factors from neighboring enterprises. 
For example, an enterprise faces a loan default, which could lead to a bad financial situation for its creditors and potentially cause bankruptcy. 
Based on these two encoders, we propose a uniform framework for capturing both enterprise intra-risk and contagion risk for bankruptcy prediction. Figure \ref{fig:mdoel-frame-work} shows the overall architecture of the proposed method.
% By doing so,
% In particular, the encoder is equipped with the two newly proposed sub-modules, i.e.  which is able to model contagion risk through two aspects, i.e. 
% the common risk factors based on hyperedges and the directed transmission risk factors from neighbors, respectively. 

% Enterprise transmission risk diffuses through SMEs knowledge graph, which remains challenge because of 
% For example, a company defaults a great deal debt, which could lead to relevant creditors' bad situation. 
% Many of this events are sure to have critical effects on a creditors' business and finally lead to its bankrupt.

% \edit{How we do? What's the main difference? Why our method works?}

In Fintech literature, especially in SMEs research, few researchers make their experimental benchmark datasets publicly available for reproduction\footnote{There are several top conference papers that did not provide their datasets for reproduction. Examples include SemiGNN \cite{wang2019semi} at ICDM 2019; HACUD \cite{hu2019cash} at AAAI 2019; ST-GNN \cite{yang2020financial} at IJCAI 2020; AMG-DP \cite{hu2020loan} at CIKM 2020; TemGNN  \cite{wang2021temporal} at SDM 2021; PC-GNN \cite{liu2021pick} at WWW 2021.}. This phenomenon, which could be related to the sensitivity of SMEs' financial data, results in a data deficiency that impedes SMEs' intelligence research \cite{yang2020financial}. 
In this study, we collect multi-sources SMEs data and build a new dataset (SMEsD), which we make publicly available. We hope the SMEsD will become a significant benchmark dataset for SMEs' bankruptcy prediction, and boost the development of financial risk research, especially SMEs bankruptcy research. 
Experimental results using the SMEsD demonstrate that our proposed model can sufficiently capture both intra-risk and contagion risk for bankruptcy prediction. 

The contributions of this work are fourfold:
\begin{itemize}
\item We conduct exploratory data analysis to demonstrate that the enterprise intelligence (i.e., enterprise basic attributes, litigation information and the EKG) affects bankruptcy risk prediction for SMEs. 

% We propose to model enterprise bankruptcy risk from the view of both intra-risk and contagion risk. Accordingly, we propose a novel uniform framework to combine them, which is a non-trivial and challenging task.

%  Under the risk fusing framework, the novel risk fusing network model is able to sufficiently mine risk information of enterprise self information and enterprise knowledge graphs.

% It is a non-trivial and challenging task to build up a uniform framework for enterprise bankruptcy prediction, taking advantage of both enterprise intra-risk and contagion risk due to the heterogeneity of the SMEs multi-source data. 
% To address this problem, we propose a novel enterprise bankruptcy prediction method by combining enterprise intra-risk and contagion risk. 

\item We propose a novel framework for inferring enterprise bankruptcy by considering both intra-risk and contagion risk. To the best of our knowledge, this is the first attempt to consider both risks simultaneously and their joint effects in bankruptcy prediction. 
\item 
Under this framework, we utilize an intra-risk encoder to derive intra-risk from an enterprise's basic intelligence. We propose a novel GNNs based contagion risk encoder that includes Hyper-GNNs and Heter-GNNs to calculate contagion risk based on hyperedges and pair-wise heterogeneous relations in the EKG.

% novel risk fusing neural network which is equipped with inner risk encoder, hierarchical transformer encoder and hierarchical hypergraph encoder, fully considering inner risk and contagion risk. 

\item We propose a new benchmark dataset (SMEsD) to evaluate the proposed method, which is also expected to further promote enterprise financial risk analysis. The empirical experiments using our dataset demonstrate that the proposed method can successfully combine enterprise intra-risk and contagion risk for bankruptcy prediction\footnote{The codes and datasets for reproduction are released on GitHub: \url{https://github.com/shaopengw/ComRisk}.}.
% We build a newly open access enterprise knowledge graph which contains abundant risk information. We conduct extensive experiments, which demonstrates the superiority of the proposed model over SOTA models.
\end{itemize}

% ACM's consolidated article template, introduced in 2017, provides a
% consistent \LaTeX\ style for use across ACM publications, and
% incorporates accessibility and metadata-extraction functionality
% necessary for future Digital Library endeavors. Numerous ACM and
% SIG-specific \LaTeX\ templates have been examined, and their unique
% features incorporated into this single new template.

% If you are new to publishing with ACM, this document is a valuable
% guide to the process of preparing your work for publication. If you
% have published with ACM before, this document provides insight and
% instruction into more recent changes to the article template.

% The ``\verb|acmart|'' document class can be used to prepare articles
% for any ACM publication --- conference or journal, and for any stage
% of publication, from review to final ``camera-ready'' copy, to the
% author's own version, with {\itshape very} few changes to the source.
\section{Exploratory Analysis}
\label{section-explortary-analysis}
In this section, we conduct an exploratory analysis of the relationship between the enterprise intelligence (i.e., enterprise basic attributes and litigation information, and the EKG), and bankruptcy risk. We first give the results for the statistical correlations and Independent Samples t Test results between the basic attributes and the lawsuit features of the enterprises and their bankruptcy status. Then, we introduce a contagion risk analysis of the EKG for bankruptcy prediction.

\subsection{Statistical Significance Analysis}
\label{section-correlation-analysis}

We collect 11,523 civil lawsuits for 3,976 Chinese SMEs from 2000 to 2021 and the basic attributes of these enterprises. 
Table \ref{tab:bankruptcy-analysis} summarizes the statistical analysis of the correlations and the Independent Samples t Test between the enterprises’ basic intelligence (i.e., enterprise basic attributes and litigation information; see Definition \ref{definition-enterprise-risk-information}) and their bankruptcy risk.
The first part in Table \ref{tab:bankruptcy-analysis} refers to enterprise basic attributes (i.e., established time, registered capital, and paid-in capital). The last four rows in Table \ref{tab:bankruptcy-analysis} concern the most significant features of lawsuits (i.e., lawsuit cause, court level, verdict, and duration of action). We present the analysis results below.

\textbf{Enterprise Attribute.}
The first part concerns enterprise basic attributes, including established time (counted by months), registered capital, and paid-in capital (counted by 10,000 yuan). From Table \ref{tab:bankruptcy-analysis}, we can find the following:

\begin{itemize}
    \item All three indicators are significantly negatively correlated with bankruptcy. The indicators of surviving enterprises are significantly higher than those of bankrupted enterprises in the t Test.
\end{itemize}

This indicates that the longer the established time, the greater the registered capital, and the greater the paid-in capital, the lower the probability of bankruptcy.

\textbf{Lawsuit Cause.}
We explore the correlation between lawsuit causes and enterprise bankruptcy. In Table \ref{tab:bankruptcy-analysis}, we find that both types of lawsuit causes (i.e., loan contract dispute and sales contract dispute) are significantly correlated with enterprise bankruptcy. 
Specifically, the correlation coefficient between the number of loan contract disputes and enterprise bankruptcy is 0.122, which is statistically significant at the 99\% level. The correlation coefficient between sales contract disputes and enterprise bankruptcy is 0.077, which is also statistically significant at the 99\% level. 
These findings confirm that bankrupted enterprises tend to have more loan contract and sales contract disputes, which is in line with intuition. Meanwhile, we can observe that the average number of loan contract disputes among the surviving enterprises is 1.80, and the number for bankrupted enterprises is 2.23. The difference between the two is significant at the 95\% level based on the t Test, which reaffirms the correlation between enterprise bankruptcy and loan contract disputes. We can obtain a similar conclusion from the statistical results for sales contract disputes. In summary, we find the following:

% \begin{itemize}
%     \item The correlation coefficient between the number of loan contract disputes and enterprise bankruptcy is significantly positive. Loan contract disputes of bankrupted enterprises are more than that of surviving enterprises significantly.
%     \item The correlation coefficient between the number of sales contract disputes and enterprise bankruptcy is significantly positive. Sales contract disputes of bankrupted enterprises are more than that of surviving enterprises significantly.
% \end{itemize}

\begin{itemize}
    \item The number of loan contract disputes and the number of sales contract disputes are both significantly positively correlated with enterprise bankruptcy.
\end{itemize}

\textbf{Court Level of Lawsuit.}
The court level of a lawsuit is another factor related to enterprise risk. There are four levels of court types (from low to high): \textit{grassroots people's court}, \textit{intermediate people's court}, \textit{higher people's court} and \textit{supreme people's court}. Most lawsuits are dealt with by the grassroots people’s court while some involving large underlying assets are brought to intermediate court. If the litigant disagrees with the verdict, it can appeal to a higher court. From Table \ref{tab:bankruptcy-analysis}, we can find the following:

% \begin{itemize}
%     \item The correlation between the number of grassroots court lawsuit and enterprise bankruptcy is significantly positive.
%     % The correlation coefficient between the number of grassroots court lawsuit and enterprise bankruptcy is significantly positive. Grassroots court lawsuits of bankrupted enterprises are more than that of surviving enterprises significantly.
%     \item The correlation between the lawsuit number of both intermediate people’s court and higher people's court and enterprise bankruptcy is significantly negative. 
%     % The correlation coefficient between the lawsuit number of both intermediate people’s court and higher people's court and enterprise bankruptcy is significantly negative. The two lawsuits of bankrupted enterprises are less than that of surviving enterprises significantly.
% \end{itemize}

\begin{itemize}
    \item The number of grassroots court lawsuits is significantly positively correlated with enterprise bankruptcy. 
    \item The lawsuit numbers of both intermediate people’s court and higher people's court are significantly negatively correlated with enterprise bankruptcy. 
    
\end{itemize}

These findings indicate that bankrupted enterprises tend to have more grassroots court lawsuits and fewer intermediate and higher court lawsuits. This could mean that being involved in a large number of grassroots court lawsuits implies that an enterprise has financial risk. Meanwhile, involvement in many high court lawsuits may reflect an enterprise’s powerful capacity to deal with lawsuits, as well as its larger business scale. The t Test confirm this conclusion.

% The difference of average number of the two types of lawsuit that bankrupt enterprises and survive enterprises 

\textbf{Verdict.}
We divide the results of lawsuits into four types according to litigant status and the verdict: plaintiff winner, plaintiff loser, defendant winner, and defendant loser. From Table \ref{tab:bankruptcy-analysis}, we can observe the following:

\begin{itemize}
    \item Enterprises that are plaintiff winners are less likely to go bankrupt (i.e., significantly negative correlation).
    \item Enterprises that are defendant losers are more prone to bankruptcy (i.e., significantly positive correlation).

\end{itemize}

The correlation coefficients of the two types of verdicts are both significant at the 99\% level, which confirms the importance of lawsuit results. The reason is that being a plaintiff winner in a lawsuit is good news for an enterprise, and being a defendant indicates risk. We can draw the same conclusion from the difference between the average number of the two types of lawsuit results for bankrupt and surviving enterprises in the t Test.

\textbf{Duration of Action.}
Referring to \cite{yin2020evaluating}, we divide duration of action (DOA) into two types: less than two years and more than two years. From Table \ref{tab:bankruptcy-analysis}, we can find the following:

\begin{itemize}
    \item The correlation between the number of lawsuits in the last two years and enterprise bankruptcy is significantly positive.
    \item The correlation between the number of lawsuits more than two years ago and enterprise bankruptcy is significantly negative.

\end{itemize}
% the correlation coefficient between number of lawsuits in two years is significant positive while that of \textbf{lawsuits before two years is significant negative}, 

These findings indicate that bankrupted enterprises tend to have had more lawsuits in the two years prior to bankruptcy. The more lawsuits, the greater the direct risk for an enterprise, especially in the case of lawsuits in the last two years. Meanwhile, having been involved in a large number of lawsuits more than two years ago implies that an enterprise has experienced many disputes but has survived. This indicates that the enterprise has a large business scale and is strong enough to face various challenges.

% \usepackage{multirow}

\subsection{Contagion-Risk Analysis}
\label{section-risk-momentum-contagion}

The contagion effect has been used to study stock movement prediction \cite{Ali2020Shared}, in which stock fluctuations are partly affected by related stocks. In this study, contagion risk means that the risk generated by an enterprise tends to diffuse through the EKG to neighboring enterprises, which is ubiquitous in real market circumstances \cite{boone2012bankruptcy,kolay2016spreading,helwege2016financial,yang2020financial}. 
Figure \ref{fig:example} shows an example of the EKG extracted from our newly generated dataset (SMEsD), from which we can find that enterprises have two types of relations: hyperedges and pair-wise heterogeneous relations. \textbf{(i)} There are three types of hyperedges in the EKG (see Definition \ref{definition-Enterprise hypergraph}): industry, area, and stakeholder, colored red, yellow, and green, respectively. For example, enterprise A, enterprise D, and enterprise E are in the same city. Then, they are influenced by the same regional policies (e.g., tax administration and economic policy) and face similar regional risks. Hence, we propose a Hyper-GNN to model such contagion risk. \textbf{(ii)} There are five types of pair-wise heterogeneous relations among enterprises and persons (see Definition \ref{definition-Enterprise heterogeneous graph}). Person 1 and enterprise F both invest in enterprise A, where the edge widths indicate distinct investment share. Person 2 and person 3 are stakeholders (e.g., manager, stockholder, or supervisor) in enterprise A and enterprise F, respectively. 
Here, we use Heter-GNNs to model this type of contagion risk.

\begin{figure*}[htb]
    \centering
    \includegraphics[width=1\textwidth]{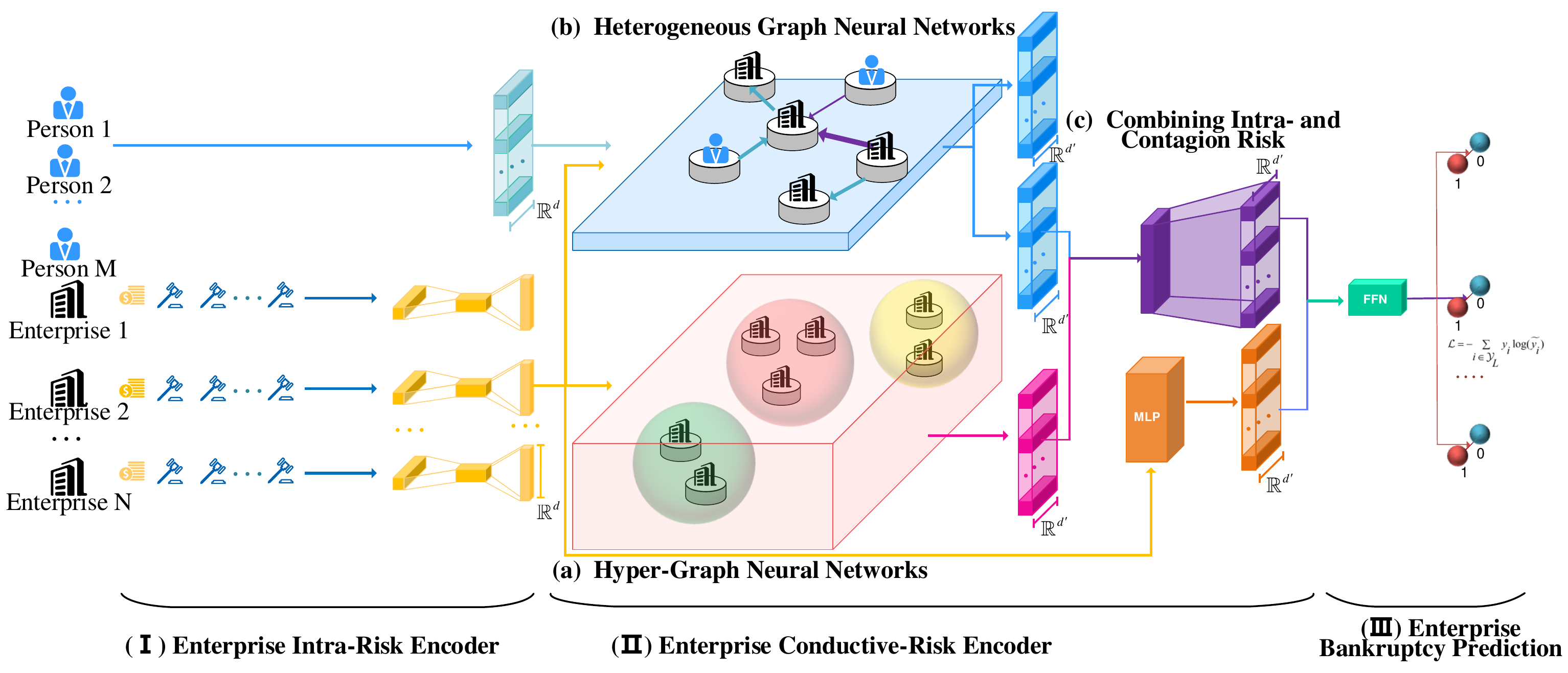}
    \caption{The overall architecture of the proposed method. \textbf{(I) Enterprise Intra-Risk Encoder} using the enterprise statistically significant features in Table \ref{tab:bankruptcy-analysis}. \textbf{(II) Enterprise Contagion-Risk Encoder} is equipped with two sub-models: \textbf{(a) Hyper-Graph Neural Networks} using enterprise hypergraph, \textbf{(b) Heterogeneous Graph Neural Networks} using enterprise heterogeneous graph, and \textbf{(c) Combining intra- and contagion risk}. \textbf{(III) Enterprise Bankruptcy Prediction}. }
    \label{fig:mdoel-frame-work}
\end{figure*}

\section{Problem Formulation}
% In this section, we present the formal definitions of data and formalize the problem of bankruptcy prediction. 
\begin{definition}
\label{definition-enterprise-risk-information}
\textbf{Enterprise basic intelligence}. 
Enterprise basic intelligence consists of two parts (i.e., the enterprise basic attributes and the enterprise litigation information as is shown in Table \ref{tab:bankruptcy-analysis}), which can be formulated as $\Acal=(\Bcal,\Jcal)$. $\Bcal=\{\b_1,\b_2,...,\b_i,...,\b_N\}$ denotes set of enterprise basic business information. $N$ denotes the number of enterprises. $\b_i=(et_i,rc_i,pc_i)$ denotes attributes for enterprise $i$, including established time, registered capital and paid-in capital. $\Jcal=\{\Jcal_1,\Jcal_2,...,\Jcal_i,...,\Jcal_N\}$ denotes enterprise litigation information of enterprises.  $\Jcal_i=\{\j_{i}^1,\j_{i}^2,...,\j_{i}^k,...,\j_{i}^K\}$ denotes lawsuit set for enterprise $i$. $K$ denotes the number of lawsuits of enterprise i. $\j_{i}^k=(lc_{i}^k,cl_{i}^k,vt_{i}^k,\Delta_{i}^k)$ denotes a specific lawsuit $k$ related to enterprise $i$, including lawsuit cause, court level of lawsuit,  verdict and time interval of action. 
% Enterprise basic intelligence consists of two parts, i.e., lawsuit data and basic business information, which can be formulated as $\Acal=(\Jcal,\Bcal)$. $\Jcal=\{\Jcal_1,\Jcal_2,...,\Jcal_n\}$ denotes lawsuit risk information of enterprises, $\Jcal_i=\{j_{i1},j_{i2},...,j_{im}\}$ denotes lawsuit set for enterprise $i$, $j_{im}=(lc_{im},cll_{im},vt_{im},doa_{im})$ denotes a specific lawsuit  $m$ related to enterprise $i$, including lawsuit cause, court level of lawsuit, verdict and duration of action. $\Bcal=\{\Bcal_1,\Bcal_2,...,\Bcal_n\}$ denotes set of enterprise basic business information. $\Bcal_n=(et_n,rc_n,pc_n)$ denotes attributes for enterprise $n$, including established time, registered capital and paid-in capital.
\end{definition}

\begin{definition}
\label{definition-Enterprise hypergraph}
\textbf{Enterprise hyper-graph}. 
An enterprise hypergraph can be defined as $\Gcal_{hyper}=(\Vcal_e, \Ecal, \Tcal_{hyper})$. Here, $\Vcal_e$ denotes the set of enterprise nodes. 
% $\Ecal=\{e_1,e_2,...,e_p\}$ 
$\Ecal=\{hp_1,hp_2,...\}$
denotes hyperedge set. $\Tcal_{hyper}=\{\Omega_1,\Omega_2,...,\Omega_M\}$ denotes hyperedge type set, and $|\Tcal_{hyper}|>1$ here. Hyperedge type map function $\psi$: $\psi(hp) \in \Tcal_{hyper}$. The relationship between enterprise nodes can be represented by an incidence matrix $\H \in \Rbb^{|\Vcal| \times |\Ecal|}$ with elements defined as:
 \begin{equation}
\H (v,hp)=\left\{
             \begin{array}{lr}
             1, & \text{if}\  v \in hp\\
             0, & \text{otherwise} \ .  \\
             
             \end{array}
\right.
\end{equation}
$v\in \Vcal_e$ denotes an enterprise node, and $hp\in \Ecal$ denotes a hyperedge. 
\end{definition}

\begin{definition}
\label{definition-Enterprise heterogeneous graph}
\textbf{Enterprise heterogeneous-graph}. 
An enterprise heterogeneous graph is defined as a connected graph $\Gcal_{hete}=(\Vcal,\Lcal,$ $\Tcal,\Rcal,\W)$. $\Vcal$ denotes the set of all nodes. 
% , which is same as that in $\Gcal_{hyper}$. 
$\Lcal$ denotes a link set. 
They are associated with two functions: (i) a node type mapping function $\varphi: \Vcal \to \Tcal$. $\Vcal = \{\Vcal_e, \Vcal_p\}$, $\Vcal_e, \Vcal_p$ denote the node set of enterprises and persons, respectively. $\Vcal_e \cap \Vcal_p = \emptyset$. Each node $v \in \Vcal$ belongs to one particular type in node type set $\Tcal: \phi(v) \in \Tcal$. 
(ii) a link class mapping function $\psi:\Lcal \to \Rcal$. $\W$ denotes edge weights.
% $\Acal$ denotes node attributes, and 
% $\forall l_1, l_2 \in \Lcal$, $\psi(l_1) \in \Rcal_\text{intra}$ and $\psi(l_2) \in \Rcal_\text{inter}$ denote the node intra-class relationships and the node inter-class relationships, respectively.
% $\Bcal\Mcal\Hcal\Gcal$ has multiple relationships, i.e., $|\Rcal_\text{inter}| > |\Tcal|-1 >0$ and $|\Rcal_\text{intra}|>1$. 
\end{definition}

\begin{problem}
\label{enterprise bankruptcy prediction}
\textbf{Enterprise bankruptcy prediction}. 
Given an enterprise multi-source data, which consists of enterprise basic intelligence $\Acal$, an enterprise heterogeneous hypergraph $\Gcal_{hyper}$ and an enterprise heterogeneous graph $\Gcal_{hete}$, we aim to determine enterprise risk, considering both intra-risk and contagion risk.
% information. 
% we aim to fuse in risk information and network structure information into enterprise node low dimension representation $\Z \in \Rcal^{|\Vcal| \times \d^\prime}$, where $\d^\prime$ is the node dimension. 
Based on enterprises' representations, we conduct bankruptcy prediction task, which can be treated as a binary classification problem. 
\end{problem}

\section{Related Work}
\subsection{Enterprise Risk Analysis}
% \textbf{risk in financial reports}

\textbf{Enterprise Intra-Risk}
In general, traditional enterprise risk analysis methods mainly consider financial indicators, such as profitability, operating efficiency, and solvency, using multivariate discriminant analyses \cite{altman1968financial,lo1986logit,lee2013multi} or machine learning methods, such as SVM  and decision trees \cite{olson2012comparative,delen2013measuring,korol2019dynamic}. For example, Erdogan et al. \cite{erdogan2021novel} propose an ensemble method utilizing SVMs as base classifiers for commercial bank bankruptcy. Many other studies use neural networks to improve prediction accuracy \cite{odom1990neural,tsai2008using}. Hosaka et al. \cite{hosaka2019bankruptcy} transform financial ratios into images using convolutional networks for bankruptcy prediction. Recently, many studies have focused on using text information, such as financial reports and conference calls, to mine enterprise intrarisk. For instance, Borochin et al. \cite{borochin2018effects} find that the tone of conference calls is negatively related to firm value uncertainty in the equity options market. Li et al. \cite{li2020maec} develop a large-scale multimodal dataset called MAEC, and their experiments demonstrate the efficiency of the dataset for volatility forecasting. Liu et al. [7] construct six pretraining tasks trained on both general and financial domain corpora, enabling them to capture financial-specific semantic information.

However, SMEs usually lack normal financial reports as well as public conference calls, which poses challenges for the analysis of SMEs. On the other hand, there are abundant risk sources such as relevant lawsuits, which are known to be significantly related to enterprise credit risk \cite{yin2020evaluating}, which have not been well utilized in previous works.

\textbf{Enterprise Contagion-Risk}
Enterprise contagion-risk is also an important part of risk analysis since no enterprise is completely independent of other companies. Some financial studies propose using interconnections between firms or assets for risk analysis \cite{eisenberg2001systemic,elsinger2006risk,fang2012simulation}. 
Eisenberg et al. \cite{eisenberg2001systemic}, for example, take interconnections among firms into consideration to study obligation-clearing mechanisms.
Elsinger et al. \cite{elsinger2006risk} propose assessing systemic financial stability using a network model of interbank loans.
% Acemoglu et al. \cite{acemoglu2015systemic} provided a framework for studying the relationship between the financial network architecture and the likelihood of systemic failures considering contagion risk and found that financial contagion exhibited a form of phase transition as inter-bank connections increase. 
% For example, \todo{in financial studies,...}
Acemoglu et al. \cite{acemoglu2015systemic} provide a framework for studying the relationship between financial network architecture and the likelihood of systemic failure considering contagion risk and find that financial contagion exhibits a form of phase transition as interbank connections increase.

% % \citet{eisenberg2001systemic} research obligation clearing mechanism considering interconnections among firms and find that unsystematic shocks to the system will lower the total value of the financial system.
% \citet{elsinger2006risk} proposed to assess systemic financial stability with a network model of interbank loan. 
% \citet{fang2012simulation} presents an interactions-based risk network, where nodes are risk and edges represent potential interactions between risks, to manage project risks.
% \citet{acemoglu2015systemic} provide a framework for studying the relationship between the financial network architecture and the likelihood of systemic failures considering contagion risk and find that financial contagion exhibits a form of phase transition as interbank connections increase. 
% However, most of previous researches explore the effect on contagion risk by simulating \cite{rahmandad2008heterogeneity,basole2014supply}, which can not be applied in real scenarios.

Most previous studies, however, explore the effects of contagion risk using simulations \cite{rahmandad2008heterogeneity,basole2014supply}, which cannot be applied to real scenarios.
\subsection{Graph Neural Networks}
% Graph Neural Networks (GNNs) utilize deep neural networks to deal with graph representation learning and witnessed a great success on various tasks on graph, such as node classification \cite{Kipf2016Semi-supervised, Velickovic2018Graph}, link prediction \cite{Abu-El-Haija2018Watch} and community detection \cite{you2019position}. GNNs also contributed a lot on traditional scenarios, such as recommendation system \cite{chang2021sequential,guo2021dual}, natural language process \cite{zhang2020every,fei2021iterative} and computer vision \cite{yang2018graph,te2018rgcnn}. We refer the readers to \cite{Chen2020Graph} for more surveys on graph neural networks. 
% GNNs shows great representation capacity on various tasks on graphs, such as node classification \cite{hu2020heterogeneous}, link prediction \cite{Abu-El-Haija2018Watch} and graph classification \cite{wu2019net}. 
Graph neural networks (GNNs) use deep neural networks to deal with graph representation learning. They have proven to be successful for various tasks on graphs, such as node classification \cite{Kipf2016Semi-supervised, Velickovic2018Graph}, link prediction \cite{Abu-El-Haija2018Watch}, and community detection \cite{you2019position}. GNNs also contribute to traditional scenarios, such as recommendation systems \cite{chang2021sequential,guo2021dual}, natural language processing \cite{zhang2020every,fei2021iterative} and computer vision \cite{yang2018graph,te2018rgcnn}. See \cite{Chen2020Graph} for more surveys of GNNs.

Enterprise interconnections naturally form a heterogeneous graph, consisting of enterprise nodes, person nodes, and the connections among them. In the fintech field, some studies use GNNs to model various risks. For example, SemiGNN \cite{wang2019semi} involves using labeled and unlabeled multiview data for fraud detection. Hu et al.
\cite{hu2020loan} model various relations, objects, the rich attributes of nodes and edges for loan default detection. CCR-GNN \cite{feng2020every} is proposed to solve the problem of corporate credit rating. Yang et al. \cite{yang2020financial} examine supply chain relationships and conduct lift prediction on a collected supply chain dataset. Kosasih et al. \cite{kosasih2021machine} pose the supply chain visibility problem as a link prediction problem via GNNs. Pan et al. \cite{pan2021heterogeneous} used a triple-layer attention network for bankruptcy prediction considering different metapath based neighbors.

% As enterprise interconnections naturally form a heterogeneous graph, consisting of enterprise nodes, person nodes and connections among them. 
% % Recently, some works concentrate on performing GNNs on enterprise risk analysis \cite{hu2019cash,wang2019semi,feng2020every} based on enterprise heterogeneous graph. instead of simulating methods.
% % For example, 
% % AMG 
% In the Fintech field, some works applied GNNs to model various risk.
% % In the Fintech field, some works use GNNs to model contagion risk \cite{hu2020loan,yang2020financial,pan2021heterogeneous}.
% For example,
% SemiGNN \cite{wang2019semi} utilized the multi-view labeled and unlabeled data for fraud detection.
% Hu et al. \cite{hu2020loan} proposed to jointly model various relations and objects as well as the rich attributes on nodes and edges for loan default detection.
% CCR-GNN \cite{feng2020every} proposed to solve the problem of Corporate Credit Rating via Graph Neural Networks.
% % ST-GNN 
% Yang et al. \cite{yang2020financial} examined mine supply chain relationship and conducted lift prediction on a collected supply chain dataset. 
% Kosasih et al. \cite{kosasih2021machine} posed the supply chain visibility problem as a link prediction problem via GNNs.
% % HAT 
% Pan et al. \cite{pan2021heterogeneous} utilized a triple-layer attention network for bankruptcy prediction considering different metapath-based neighbors. 

Hypergraphs have shown a strong capacity to model higher-order relationships, which have been used in many areas, such as social recommendation \cite{bu2010music,yu2021self} and computer vision \cite{wu2020adahgnn,liu2020semi}. With regard to enterprise risk modeling, there is a large number of hyperedges among enterprises and related persons, which is suitable for using hypergraphs. Few studies, however, have applied hypergraph neural networks in this area.

As discussed previously, few researchers have considered both intra-risk and contagion risk simultaneously in relation to bankruptcy prediction. Further, most fail to sufficiently mine risk information because of complex risk sources and relationships. 
Meanwhile, few studies provide open access data for other researchers, which restricts the development of risk analysis research in areas such as bankruptcy prediction and default prediction.

\section{Methodology}
\label{section-method}
In this section, we introduce the overall architecture of the proposed method, as shown in Figure \ref{fig:mdoel-frame-work}. The proposed model consists of three main parts: {\textbf{(I)} the enterprise intrarisk encoder}, which uses statistically significant enterprise features (Table \ref{tab:bankruptcy-analysis}); {\textbf{(II)} Enterprise Contagion Risk Encoder}, which consists of two submodules: (a) Hyper-GNNs, using enterprise hypergraphs, and (b) Heter-GNNs, using enterprise heterogeneous graphs; and (c) Combining intra- and contagion risk. {\textbf{(III)} Enterprise Bankruptcy Prediction}. Different from previous work, we take advantage of a hierarchical mechanism for both Hyper-GNNs and Heter-GNNs to utilize complex heterogeneous hyperedges and relationships. We provide the details below.

% In this section, we introduce the overall architecture of the proposed method, as shown in Figure \ref{fig:mdoel-frame-work}. The proposed model consists of three significant parts: {\textbf{(I)} Enterprise Intra-Risk Encoder} using the enterprise statistically significant features in Table \ref{tab:bankruptcy-analysis}. {\textbf{(II)} Enterprise Contagion Risk Encoder} consists of two sub-modules: (a) Hyper-Graph Neural Networks (Hyper-GNNs) using enterprise hypergraph, {(b) Heterogeneous Graph Neural Networks (Heter-GNNs)} using enterprise heterogeneous graph, and (c) Combining intra- and contagion risk. {\textbf{(III)} Enterprise Bankruptcy Prediction}. Different from previous works, we take advantage of hierarchical mechanism for both Hyper-GNNs and Heter-GNNs to sufficiently utilize multiplex heterogeneous hyperedges and heterogeneous relations. Next, we give the details of them. 

% , Risk fusing network consists of three parts, i.e., inner risk encoder, hierarchical transformer encoder and hierarchical hypergraph encoder. 

\subsection{Enterprise Intra-Risk Encoder}
\label{Enterprise Intra-Risk Encoder}
The enterprise intra-risk encoder aims to learn enterprise self-risk
embedding using enterprise basic intelligence (i.e., enterprise basic attributes and enterprise litigation information), which is formally given in Definition \ref{definition-enterprise-risk-information}.

% The enterprise intra-risk encoder aims to learn enterprise self risk embedding using enterprise basic intelligence, i.e. enterprise basic attributes and enterprise litigation information, which is formally given in Definition \ref{definition-enterprise-risk-information}. 
% through Glorot \cite{glorot2010understanding}.

\begin{figure}[htb]
    \centering
    \includegraphics[width=0.44\textwidth]{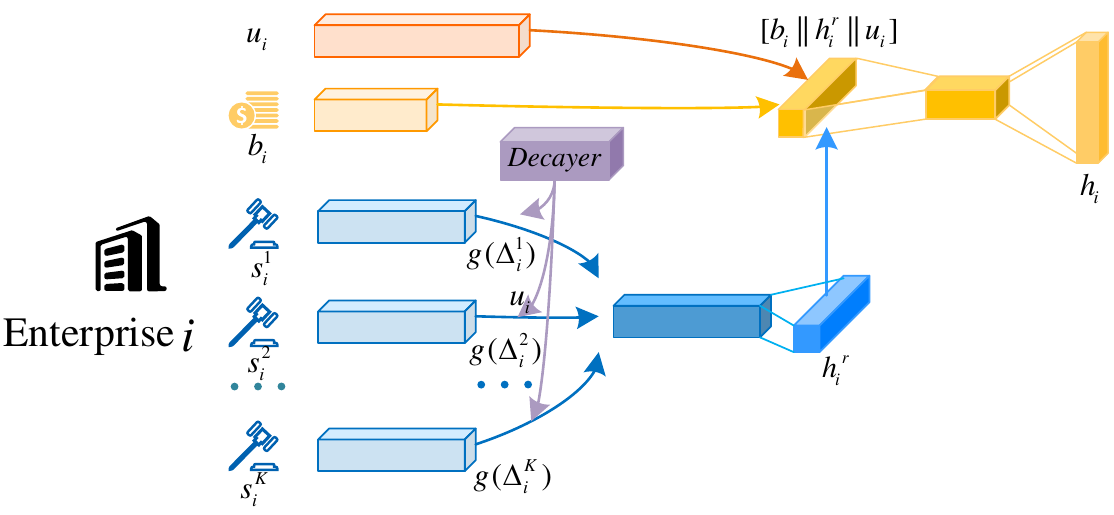}
    \caption{Enterprise Intra-Risk Encoder. }
    \label{fig:intra-risk}
\end{figure}

As Figure \ref{fig:intra-risk}, first, for each enterprise node $v_i \in \Vcal_e$, we use $\b_i \in \Rbb^{\hat{d}}$ in Definition \ref{definition-enterprise-risk-information} as the basic attribute features. Second, the lawsuit event $\j_i^k$ of enterprise $i$ contains four significance attributes (i.e., lawsuit cause, court level, verdict, and DOA), as described in Section \ref{section-correlation-analysis}. For the first three attributes, we map each into latent spaces and then concatenate them to obtain lawsuit representation $\s_i^k \in \Rbb^{\tilde{d}}$. Referring to \cite{ma2020streaming}, we use a time decay function \textit{Decayer} to weight each lawsuit representation to make better use of time information in lawsuit events. Specifically, we calculate the time interval $\Delta_i^k$ between the time of occurrence of each lawsuit and the enterprise’s observation time. For bankrupted enterprises, the observation time is set as the time of bankruptcy while for surviving enterprises, it is set as the present.

% As Fig. \ref{fig:intra-risk} shows, firstly, for each enterprise node $v_i \in \Vcal_e$, we use $\b_i \in \Rbb^{\hat{d}}$ in Definition \ref{definition-enterprise-risk-information} as its basic attribution features. 
% % we map node $v_i$'s business attributes $a_i \in \Bcal_i$ into corresponding vector spaces and concatenate them to get enterprise business attribute embedding $\p_i$. 
% Secondly, the lawsuit event $\j_i^k$ of enterprise $i$ contains four significance attributes, i.e., lawsuit cause, court level of lawsuit, verdict and duration of action, as analysis in Section \ref{section-correlation-analysis}. For the first three attributes, we map each of them into latent spaces and then concatenate them to get lawsuit representation $\s_i^k \in \Rbb^{\tilde{d}}$.  Inspired by \cite{ma2020streaming}, we utilize a time decay function \textit{Decayer} to weight each lawsuit representation for better making use of time information in relevant lawsuit events. Specifically, we calculate time interval $\Delta_i^k$ between the happened time of each related lawsuit and the enterprise's observation time. For bankrupted enterprises, the observation time is set as bankruptcy time, while for surviving enterprises the observation time is set as the present time. 
\begin{equation}
\begin{array}{l}
\begin{aligned}
    \label{equation-time-decayer}
     g(\Delta_i^k)=\frac{1}{1+ w \cdot \Delta_i^k}\ ,\\
\end{aligned}
\end{array}
\end{equation}

Because lawsuits in the past two years play an important role in enterprise risk prediction \cite{yin2020evaluating}, we assign a lower $w$ when performing time weight decay for lawsuits in the last two years. 

% As the most recent two years' lawsuits play an important role for enterprise risk prediction \cite{yin2020evaluating}, we assign lower $w$ when performing time weight decay for the recent two years' lawsuits. 
% We divide lawsuits into $L$ periods $\{ T_1,T_2,...,T_L\}$. Then, we sum lawsuit information in the same time period based on decayed lawsuit representations.
% Afterwards, we utilize LSTM  \cite{hochreiter1997long} to aggregate lawsuit information from different time periods as follows:
% \begin{equation}
% \begin{array}{l}
% \begin{aligned}
%     \label{equation-lstm}
%      \C_{it}^{(l)},\ \h_{it}=\textit{LSTM}(\C_{i(t-1)}^{(l)},\ \h_{it}^{(l)})\ , \\
% \end{aligned}
% \end{array}
% \end{equation}
Then, we aggregate lawsuit information from different time periods as follow:
\begin{equation}
\begin{array}{l}
\begin{aligned}
    \label{equation-lawsuit-weight-pooling}
     \h_{i}^{r}=\sum \limits_{k \in K_i} \W_{risk} \ g(\Delta_i^k) \cdot \s_i^{k}\ ,\\
\end{aligned}
\end{array}
\end{equation}
where $\W_{risk} \in \Rbb^{\tilde{d} \times d} $ is a trainable matrix, $\h_{i}^{r}$ is the aggregated lawsuit information of company $i$.
% \begin{equation}
% \begin{array}{l}
% \begin{aligned}
%     \label{equation-lstm-ft}
%      \f_t=\sigma\Big(\W_f\cdot [\c_{i(t-1)}^{(l)}, \h_{it}^{(l)}]+\b_f\Big) \ ,\\
% \end{aligned}
% \end{array}
% \end{equation}
% \begin{equation}
% \begin{array}{l}
% \begin{aligned}
%     \label{equation-lstm-it}
%      \i_t=\sigma\Big(\W_i\cdot[\c_{i(t-1)}^{(l)}, \h_{it}^{(l)}]+\b_i\Big) \ ,\\
% \end{aligned}
% \end{array}
% \end{equation}
% \begin{equation}
% \begin{array}{l}
% \begin{aligned}
%     \label{equation-lstm-ot}
%      \o_t=\sigma\Big(\W_o\cdot[\c_{i(t-1)}^{(l)}, \h_{it}^{(l)}]+\b_o\Big) \ ,\\
% \end{aligned}
% \end{array}
% \end{equation}
% \begin{equation}
% \begin{array}{l}
% \begin{aligned}
%     \label{equation-lstm-ct-t}
%      \tilde{\C}_{it}=\textit{tanh}\Big(\W_c\cdot[\c_{i(t-1)}^{(l)}, \h_{it}^{(l)}]+\b_c\Big) \ ,\\
% \end{aligned}
% \end{array}
% \end{equation}
% \begin{equation}
% \begin{array}{l}
% \begin{aligned}
%     \label{equation-lstm-ct}
%      \C_{it}=\f_t\ast \C_{i(t-1)}+\i_t\ast\tilde{\C}_{it} \ ,\\
% \end{aligned}
% \end{array}
% \end{equation}
% \begin{equation}
% \begin{array}{l}
% \begin{aligned}
%     \label{equation-lstm-ht}
%      \h_{it}=\o_t\ast \textit{tanh}\Big(\C_{it}\Big) \ , \\
% \end{aligned}
% \end{array}
% \end{equation}
% For convenience, we summarize the procedure of the LSTM unit
% in eq. \ref{equation-lstm-ft} to eq. \ref{equation-lstm-ht} as following:
% \begin{equation}
% \begin{array}{l}
% \begin{aligned}
%     \label{equation-lstm-sum}
%      \h_i^{(r)}=\sum \limits_{t=1}^{L} \h_{it} \ .\\
% \end{aligned}
% \end{array}
% \end{equation}

We also generate a pre-trained embedding $\u_i \in \Rbb^{\overline{d}}$ for enterprise $i$ as a supplement embedding.
% Thirdly, we also randomly generate an embedding $\u_i$ for enterprise $i$ based on standard normal distribution as a supplement embedding, since we believe its hidden risk is always unknown. 
% Finally, we concatenate total the basic attribution features, the litigation  embedding and supplement embedding, and project it into new latent space as follow:
Finally, we concatenate the basic attribution features, litigation embedding, and supplement embedding and project it into a new latent space as follow:

 \begin{equation}
\begin{array}{l}
\begin{aligned}
    \label{equation-risk-info}
     \h_i=\W_e \cdot [\b_i || \h_i^{r} || \u_i ]\ .\\
\end{aligned}
\end{array}
\end{equation}
$\h_i$ denotes the output of intra-risk representation of the enterprise $i$.
% information of enterprise node $i$ fusing basic attributes and the total litigation information. 
$||$ denotes the concatenation operation. $\W_e \in \Rbb^{(\hat{d}+d+\overline{d}) \times d}$ is a trainable matrix. 
% Figure \ref{fig:intra-risk} shows the detailed architecture of the intra-Risk Encoder. 

% Then we combine enterprise initial embedding with the business attribute embedding as follows:
% \begin{equation}
% \begin{array}{l}
% \begin{aligned}
%     \label{equation-concatenate-u-z}
%      \h_i^{(b)}=\u_i||\p_i \ ,\\
% \end{aligned}
% \end{array}
% \end{equation}
% where $\h_i^{(b)}$ denotes enterprise node $i$'s combined representation, 

% Given a set of enterprises $\Vcal_e$,
% for each node $v\in \{\Vcal_e,\Vcal_p\}$, 

\subsection{Enterprise Contagion Risk Encoder}
\subsubsection{Hyper-Graph Neural Networks}
Hypergraphs play an important role in bankruptcy prediction, as the hyperedges reflect common factors that enterprises face. Thus, it is natural to utilize hypergraphs to capture common risk information, such as industry development recession, regional economic policy changes, and guarantee risk caused by the same stakeholders. 

\begin{figure}[htb]
    \centering
    \includegraphics[width=0.4\textwidth]{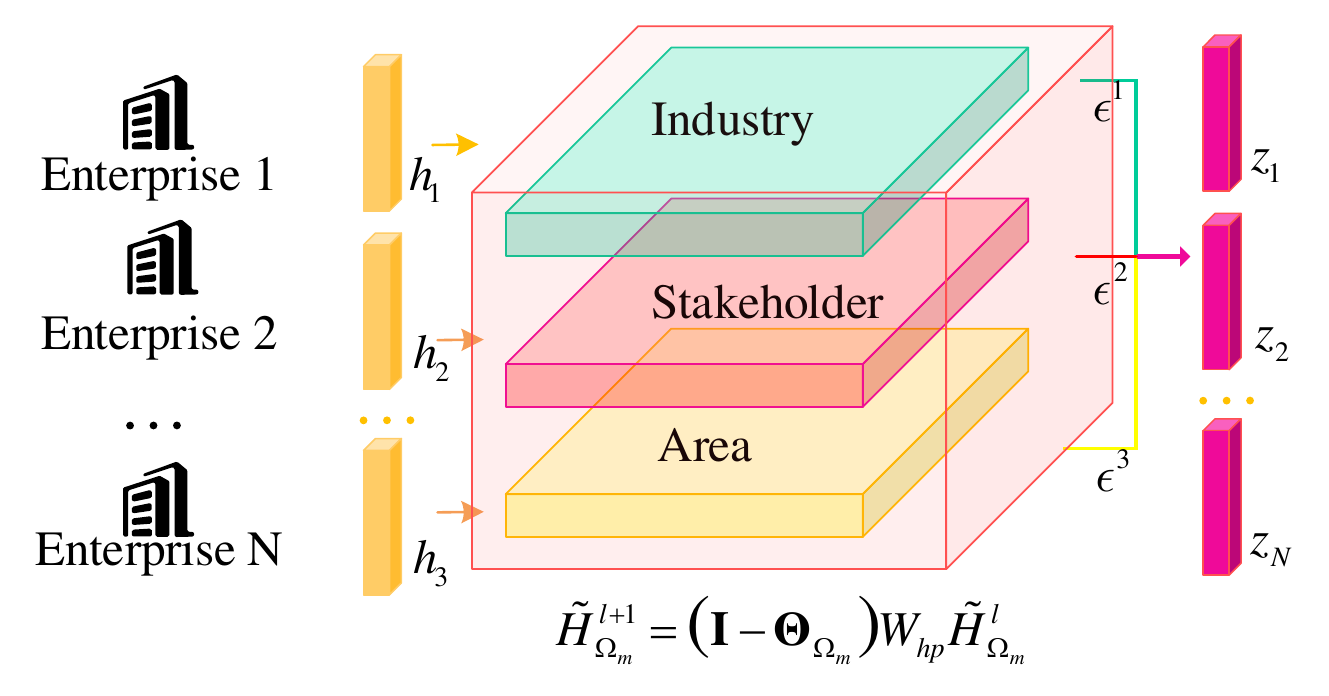}
    \caption{Hyper-Graph Neural Networks. }
    \label{fig:hyperGNNs}
\end{figure}

As shown in Figure \ref{fig:hyperGNNs}, because different types of hyperedges contribute to node representation at different levels, we assign different weights to them when aggregating node representations. Specifically, following Feng et al. \cite{feng2019hypergraph}, we first calculate the hypergraph convolution module as follow:
\begin{equation}
    \label{equation-hypergraph-hyperedge-theta}
    \boldsymbol{\Theta}_{\Omega_m}=\D_v^{-1/2}\H_{\Omega_m}\W\D_e^{-1}\H_{\Omega_m}^\top \D_v^{-1/2} \ ,
\end{equation}
where $\boldsymbol{\Theta}_{\Omega_m} \in \Rbb^{|\Vcal_e|\times |\Vcal_e|}$ denotes the convolution module. $\D_v$ is the enterprise node degree matrix. $\H_{\Omega_m}$ denotes the incident matrix of the hypergraph type $\Omega_m$. $\W$ is the node weight matrix. We set it as an identity matrix, which means all weights are equal. $\D_e$ denotes the hyperedge degree matrix. Afterwards, we conduct hypergraph convolution under the hypergraph type $\Omega_m$ as follow:
\begin{equation}
    \label{equation-hypergraph-hyperedge-convolution}
    \widetilde{\H}_{\Omega_m}^{l+1}=  \Big(\boldsymbol{I}-\boldsymbol{\Theta}_{\Omega_m}\Big)\W_{hp}\widetilde{\H}_{\Omega_m}^{l},
\end{equation}
where $\widetilde{\H}_{\Omega_m}^{l+1}$ denotes the learned representations under the hypergraph type ${\Omega_m}$ of layer $l+1$, $\boldsymbol{I}-\boldsymbol{\Theta}_{\Omega_m}$ denotes the hypergraph laplacian, $\W_{hp} \in \Rbb^{\d \times \d^\prime}$ is a trainable matrix, which is shared for different types of hypergraphs. 
Then we aggregate the different types of hypergraph convolution representations as follow:

% \begin{equation}
% \label{equation-hypergraph-hyperedge-attention}
%   f_i^{\Omega_m}=\tilde{\q}^\top h_i^{\Omega_m},
% \end{equation}

% \begin{equation}
% \label{equation-hypergraph-hyperedge-softmax}
%     \epsilon_{i}^{\Omega_m} = \text{Softmax}_m (f_{i}^{\Omega_m}) = \frac{\exp{(f_{i}^{\Omega_m})}}{\sum\limits_{\Omega_n \in \Rcal}\exp{(f_{i}^{\Omega_n})}} \ ,
% \end{equation}

% \begin{equation}
% \label{equation-hypergraph-importance}
%   f_i^{\Omega_m}=\tilde{\q}^\top h_i^{\Omega_m},
% \end{equation}

\begin{equation}
    \label{equation-hypergraph-hyperedge-aggregation}
    \z_i  = \sum_{\Omega_m \in \Tcal_{hyper}} \epsilon^{\Omega_m }\cdot \widetilde{\h}_i^{\Omega_m}, 
\end{equation}
where $\z_i \in \Rbb^{d^\prime}$ is the learned hypergraph comprehensive representation of enterprise $i$, and $\epsilon^{\Omega_m }$ is a trainable parameter, which denotes the importance of hypergraph ${\Omega_m }$ for all enterprise nodes. 
% Figure \ref{fig:hyperGNNs} shows the architecture of Hyper-GNNs. 

\subsubsection{Heterogeneous Graph Neural Networks}
\begin{figure}[htb]
    \centering
    \includegraphics[width=0.4\textwidth]{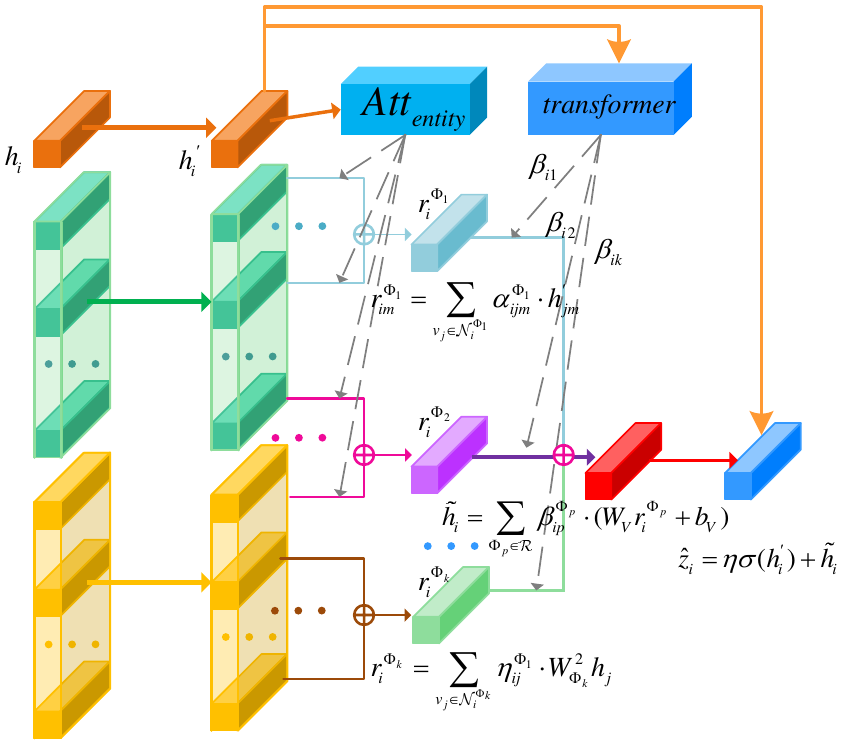}
    \caption{Heterogeneous Graph Neural Networks. }
    \label{fig:heteGNNs}
\end{figure}

We propose the Heter-GNNs to sufficiently make use of multiplex interactions among enterprises and persons. Specifically, we first aggregate entity level information and then relationship level in a hierarchical mechanism as shown in Figure \ref{fig:heteGNNs}.

 We initialize the person node representations the same as for enterprises in Section \ref{Enterprise Intra-Risk Encoder}. Then we perform transformation based on node type to project enterprise node and person representation to same latent space as follow:
 \begin{equation}
\begin{array}{l}
\begin{aligned}
    \label{equation-hierarchical-HIN-projection}
     \h_i^{\prime}=\text{Norm}(\W_{\phi(v_i)} \h_i)\ ,\\
\end{aligned}
\end{array}
\end{equation}
 where $\W_{\phi(v_i)} \in \Rbb^{d \times d^{\prime}}$ is a node type specific trainable weight matrix. $\h_i \in \Rbb^{d}$ and $\h_i^{\prime} \in \Rbb^{d^\prime}$ are the original and transformed node representations, respectively. $\text{Norm}$ denotes Batch Normalization operation \cite{ioffe2015batch}.
%  Noted that the original embedding  $\d$ is different for enterprise and person node, while the transformed dimension $\d^{\prime}$ is same for the two types of node.
Then we conduct entity level aggregation. For weighted edges, such as \textit{holder\_investor}, we directly set the ratio of contribution capital as the edge weight. For unweighted relations, we use the attention mechanism to assign weights for node $v_i$'s neighbors' representation as follows:
\begin{equation}
\begin{array}{l}
\begin{aligned}
    \label{equation-hierarchical-HIN-entity-level-attention}
     \e_{ij}^{\Phi_k}=  \text{Att}_{entity}(\h_i^{\prime},\h_j^{\prime};\Phi_k)
     = \text{LeakyRelu}(\W_{\Phi_k}^1 \cdot [ \h_i^{\prime} \| \h_j^{\prime}]) \ ,
\end{aligned}
\end{array}
\end{equation}
where $\e_{ij}^{\Phi_k}$ is the learned importance of node $i$'s neighbor $j$ under relationship $\Phi_k$, $\W_{\Phi_k}^1 \in \Rbb^{2d^\prime \times d^\prime}$ is a trainable matrix, and \text{LeakyRelu} is an activation function. 
To make the weights comparable, we utilize \textit{Softmax} function to normalize weights across all choices of j as follows:
\begin{equation}
\begin{array}{l}
\begin{aligned}
\label{equation-hierarchical-HIN-entity-level-softmax}
    \alpha_{ijm}^{\Phi_k} = \textit{Softmax}_j (e_{ijm}^{\Phi_k}) = \frac{\exp{(e_{ijm}^{\Phi_k}})}{\sum\limits_{v_p \in \Ncal^{\Phi_k}(v_i) }\exp{(e_{ipm}^{\Phi_k})}} \ ,
\end{aligned}
\end{array}
\end{equation}

\begin{equation}
\begin{array}{l}
\begin{aligned}
    \label{equation-hierarchical-HIN-entity-level-aggregation}
     r_{im}^{\Phi_k}  = \sum_{v_j \in \Ncal_i^{\Phi_k}} \alpha_{ijm}^{\Phi_k }\cdot h_{jm}^\prime\ ,\\
\end{aligned}
\end{array}
\end{equation}
where $r_{im}^{\Phi_k}$ is the $m$-th element of the aggregated ${\Phi_k}$ unweighted relationship representation for node $v_i$. $\alpha_{ijm}^{\Phi_k}$ is the $m$-th dimension of the normalized importance of node $j$ related to node $i$ under the unweighted relationship $\Phi_k$, $\Ncal_i^{\Phi_k}$ denotes node $i$'s neighbors under unweighted relationship $\Phi_k$. 
For weighted edges, we implement node level aggregation as follows:
\begin{equation}
\begin{array}{l}
\begin{aligned}
\label{equation-hierarchical-HIN-entity-level-softmax}
    \eta_{ij}^{\Phi_k} = \textit{Softmax}_j (w_{ij}^{\Phi_k}) = \frac{\exp{(w_{ij}^{\Phi_k}})}{\sum\limits_{v_p \in \Ncal^{\Phi_k}(v_i) }\exp{(w_{ip}^{\Phi_k})}} \ ,
\end{aligned}
\end{array}
\end{equation}

\begin{equation}
\begin{array}{l}
\begin{aligned}
    \label{equation-hierarchical-HIN-entity-level-aggregation}
     \r_i^{\Phi_k}  = \sum_{v_j \in \Ncal_i^{\Phi_k}} \eta_{ij}^{\Phi_k }\cdot \W_{\Phi_k}^2\bm{\h}_j \ ,\\
\end{aligned}
\end{array}
\end{equation}
where $\eta_{ij}^{\Phi_k}$ denotes the normalized importance that node $j$ has for node $i$ under weighted relation, and $w_{ij}^{\Phi_k}$ denotes original edge weight between node $i$ and node $j$ (e.g., such as contribution capital). $\W_{\Phi_k}^2 \in \Rbb^{d^\prime \times d^\prime}$ is a trainable matrix. $\r_i^{\Phi_k}$ denotes the learned aggregated representation of node $i$'s neighbors under the weighted relationship $\Phi_k$.

To fully capture the risk information implied in different relationships, we use transformer based attention mechanism:
% as follows:
\begin{equation}
\begin{array}{c}
\begin{aligned}
    \label{equation-hierarchical-HIN-relation-level-attention}
     g_{ik}=\k_i^\top\q_i\cdot \frac{\mu_{\Phi_k}}{\sqrt{d^\prime}},\\
     \q_i=\W_Q^{\Phi_k}\h_{i}^{\prime}+\b_Q^{\Phi_k} \ ,\\
     \k_i=\W_K^{\Phi_k}\r_{i}^{\Phi_k}+\b_K^{\Phi_k}\ ,\\
\end{aligned}
\end{array}
\end{equation}
where $g_{ik}$ denotes the relationship level importance that relation $\Phi_k$ has for node $i$, and $\W_Q^{\Phi_k}, \W_K^{\Phi_k} \in \Rbb^{d^\prime \times d^\prime}$ are trainable matrices in relationship $\Phi_k$, $\b_Q^{\Phi_k}, \b_K^{\Phi_k} \in \Rbb^{d^\prime}$ are trainable parameters in relationship $\Phi_k$, $\mu_{\Phi_k}$ is a trainable parameter used to adjust the scale of learned importance, which is relationship type specific. Similarly, we utilize the \textit{Softmax} function to normalize learned attention and aggregate relation level representations as follows:
\begin{equation}
\begin{array}{l}
\begin{aligned}
    \label{equation-hierarchical-HIN-relation-level-softmax}
     \beta_{ik}=\textit{Softmax}_k(g_{ik})=\frac{\exp(g_{ik})}{\sum \limits_{\Phi_p\in \Rcal}\exp(g_{ip})} \ ,\\
\end{aligned}
\end{array}
\end{equation} 

\begin{equation}
    \label{equation-hierarchical-HIN-relation-level-aggregation}
    \widetilde{\h_i} = \sum_{\Phi_p \in \Rcal} \beta_{ip}^{\Phi_p}\cdot (\W_V \r_i^{\Phi_p}\ +\b_V).
\end{equation}

where $\beta_{ik}$ denotes the normalized importance of relationship $\Phi_k$ for node $i$, $\W_V \in \Rbb^{d^\prime \times d^\prime}$ and $b_V \in \times \Rbb^{d^\prime}$ are trainable parameters. $\widetilde{\h_i}$ is the learned aggregated risk information for node $i$.  
Next, we use the residual connection to get the final risk information of the heterogeneous graph as follow:
\begin{equation}
    \label{equation-hetegnns-resideal-connection}
    \bm{\hat{z}}_i  =\eta \sigma(\h_i^\prime)+ \widetilde{\h_i} ,
\end{equation}
where $\eta$ is the learned weight to balance the aggregated risk information and nodes' initial risk information, $\sigma$ is the GELU \cite{hendrycks2016gaussian} activation function and $\bm{\hat{z}}_i$ is the final risk information of node $i$.

\subsubsection{Combining Intra- and Contagion Risk}
We sum the propagated risk from the Hyper-GNNs and Heter-GNNs to get the contagion-risk as follow:
\begin{equation}
    \label{equation-contagion-risk}
    \bm{\z}_i^{cont}  = \W^{cont} \cdot (\z_i + \bm{\hat{z}}_i) ,
\end{equation}
where $\W^{cont} \in \Rbb^{d^\prime \times d^\prime }$ is a trainable matrix and $\bm{\z}_i^{cont}$ is the learned contagion risk.

Then, we combine the node intra-risk and contagion-risk information as follow:

\begin{equation}
    \label{equation-residual-connection}
    \bar{\bm{\z}}_i  = \lambda \sigma(\bm{\z}_i^{cont}) + (1-\lambda) \text{MLP} (\widetilde{\h}_i) \ . 
\end{equation}
where $\bar{\bm{\z}}_i$ is the final representation of node $i$, and $\lambda$ is a trainable parameter to balance contagion risk and intra-risk. $\sigma$ is an activation function, we choose GELU here. \text{MLP} is a two-layer multilayer perception with the ReLU \cite{shang2016understanding} activation function in it.
% Figure \ref{fig:heteGNNs} shows the architecture of Heter-GNNs.

\subsection{Optimization}
We sum the learned representations of Hyper-GNNs and Heter-GNNs and utilize a fully connected layer to transform learned node representations for bankruptcy prediction, as in Figure \ref{fig:mdoel-frame-work} (III).

\begin{equation}
\label{equation-final-predict}
    \tilde{y_i}=\textit{Softmax}\Big(\W_p \bar{\bm{\z}}_i+\b_p \Big) \ ,
\end{equation}
where $\W_p$ is a trainable matrix and $\b_p$ is the bias vector. 
Finally we train the model by minimizing cross-entropy loss. 
\begin{equation}
\label{loss-function}
    \Lcal=-\sum\limits_{i \in\Ycal_{L}} y_i\log(\tilde{y_i}) \ .
\end{equation}
where $\Ycal_{L}$ is the set of labeled nodes. $y_i$ and $\tilde{y_i}$ are the ground truth and the predicted bankruptcy probability for node $i$, respectively.

\section{Experiments}
\subsection{Experimental Settings}
\subsubsection{Datasets}
% To examine the performance of the proposed model on bankruptcy prediction, we manually collect and pre-process a real world SMEs dataset, named \textbf{SMEsD}. To the best of our knowledge, this dataset is the largest multi-mode bankruptcy prediction dataset that contains abundant multi-dimension information. 
% % We firstly choose 889 bankruptcy enterprises in Zhejiang province as the seed enterprise and then collect their related enterprises and associated persons to build up SMEsD.
% The SMEsD consists of 3,976 SMEs and related persons in China from 2014 to 2021, which constitutes a multiplex enterprise knowledge graph. All enterprises are associated with their basic business information and lawsuit events spanning from 2000 to 2021. Specifically, the enterprise business information includes registered capital,  paid-in capital and established time. Each lawsuit consists of the associated plaintiff, defendant, subjects, court level, result and timestamp.

To examine the performance of the proposed model for bankruptcy prediction, we manually collect and preprocess a real-world SME dataset, which we call \textbf{SMEsD}. To the best of our knowledge, this dataset is the largest multimode bankruptcy prediction dataset, and it contains abundant multidimensional information. SMEsD consists of 3,976 SMEs and related persons in China from 2014 to 2021, constituting a complex EKG. All enterprises are associated with their basic business information and lawsuit events spanning 2000–2021. Specifically, enterprise business information includes registered capital, paid-in capital, and established time. Each lawsuit consists of the associated plaintiff, defendant, subjects, court level, result, and timestamp.

% Table \ref{tab:Statistics} gives the statistics of the SMEsD. The dataset contains two types of nodes, i.e., enterprise and person. For the enterprise heterogeneous graph, there exist 5 types of relationships between enterprises and persons. 
% The \textit{holder\_investment} relationship is weighed by the contribution capital and other edges are unweighted. 
% % Besides, the $loan$ and $deal$ edges are constructed through lawsuits if enterprises have such two types of disputes. 
% For the hypergraph, there exist three types of edges, i.e. industry, area and stakeholder. We split SMEsD into training set, validation set and testing set across the bankruptcy time of seed enterprises.

Table \ref{tab:Statistics} presents the statistics of the SMEsD. The dataset contains two types of nodes: enterprise and person. For the enterprise heterogeneous graph, there are five types of relationships between enterprises and persons. The \textit{holder\_investor} relationship is weighed by the contribution capital, and the other edges are unweighted. For the hypergraph, there are three types of edges: industry, area, and stakeholder. We split SMEsD into a training set, validation set, and testing set across the bankruptcy time.

% \usepackage{multirow}

% \usepackage{multirow}

% \usepackage{multirow}
\begin{table}[]
\caption{Statistics of the SMEsD}
\label{tab:Statistics}
\begin{tabular}{cl|c|c|c}
\toprule
\multicolumn{2}{c|}{\textbf{SMEsD}}                                                & \textbf{Train} & \textbf{Validation} & \textbf{Testing} \\ \midrule
\multicolumn{1}{c|}{\multirow{2}{*}{Node}}   & \textit{\#company}            & 2848  & 741   & 505  \\
\multicolumn{1}{c|}{}                        & \textit{\#person}             & 1752  & 367   & 322  \\ \midrule
\multicolumn{1}{c|}{\multirow{9}{*}{HeteG}}  & \textit{\#manager}            & 2658   & 724   & 562   \\
\multicolumn{1}{c|}{}                        & \textit{\#shareholder}        & 4002   & 1016   & 704  \\
\multicolumn{1}{c|}{}                        & \textit{\#other stakeholder}  & 4426  & 1028   & 948  \\
\multicolumn{1}{c|}{}                        & \textit{\#holder\_investor} & 6626  & 1574   & 1208  \\
\multicolumn{1}{c|}{}                        & \textit{\#branch}             & 594   & 98    & 76   \\
% \multicolumn{1}{c|}{}                        & \textit{\#loan}               & 198    & 96    & 44   \\
% \multicolumn{1}{c|}{}                        & \textit{\#deal}               & 26    & 4     & 10    \\
 \midrule
\multicolumn{1}{c|}{\multirow{3}{*}{HyperG}} & \textit{\#industry}           & 108    & 82    & 68   \\
\multicolumn{1}{c|}{}                        & \textit{\#area}               & 153   & 61    & 61   \\
\multicolumn{1}{c|}{}                        & \textit{\#stakeholder}        & 756  & 164   & 152   \\ \midrule
\multicolumn{1}{c|}{\multirow{2}{*}{Label}}  & \textit{\#bankrupt}           & 1621  & 354   & 318  \\
\multicolumn{1}{c|}{}                        & \textit{\#survive}            & 1195  & 367   & 173  \\ \bottomrule
\end{tabular}
\end{table}

\subsubsection{Baselines} 
% To measure the effectiveness of our method, we compare the proposed model with four types of state-of-the-art (SOTA) methods: \textbf{(1)} the conventional machine learning based method that only consider enterprise lawsuit information in the view of lawsuit attributes' frequency and basic business information; \textbf{(2)} the hypergraph neural networks based methods that take high order relationships among enterprises into consideration, which is able to detect common risk that enterprises are confronted with. \textbf{(3)} the homogeneous GNNs based methods that utilize abundantly connections among enterprises, which is able to capture contagion risk; \textbf{(4)} the heterogeneous GNNs based methods that is able to distinguish multiplex relationships in EKG.

To measure the effectiveness of our method, we compare the proposed model with four types of state-of-the-art (SOTA) methods: \textbf{(1)} the conventional machine learning based method that only considers enterprise lawsuit information including lawsuit attribute frequency and basic business information; \textbf{(2)} hypergraph neural networks based methods that take high-order relationships among enterprises into consideration; this can detect the common risks enterprises face; \textbf{(3)} homogeneous GNNs based methods that use abundant connections among enterprises, which can capture contagion risk; and \textbf{(4)} heterogeneous GNNs based methods that can distinguish complex relationships in an EKG.

\textit{Conventional Machine Learning (ML) Based Methods}
\begin{itemize}
    \item Logistic Regression (\textbf{LR}) \cite{hosmer2013applied}: a well known method applied in machine learning, social science and biometrics when explained variables are discrete.
    \item Support Vector Machine (\textbf{SVM}) \cite{suykens1999least}: a model utilized support vectors to divide vector spaces into different classes.
    \item Gradient Boosting Decision Tree (\textbf{GBDT}) \cite{friedman2001greedy}: a classic tree classification model of conventional machine learning.
\end{itemize}

\textit{Hypergraph Neural Networks (HyperG) Based Methods} 

\begin{itemize}
    \item Hypergraph Neural Networks (\textbf{HGNN}) \cite{feng2019hypergraph}: a model proposed to utilize high-order relationship information in graphs. 
    \item Hypergraph Wavelet Neural Network (\textbf{HWNN}) \cite{sun2021heterogeneous}: a newly proposed model which makes use of wavelet basis instead of Fourier basis to perform localized hypergraph convolution. 
\end{itemize}

% \item Random Walk based methods
\textit{Homogeneous GNNs (HomoG) Based Methods}
\begin{itemize}
    \item Graph Convolutional Networks (\textbf{GCN}) \cite{Kipf2017Semi-supervised}: a popular model which averages neighbors' information during the message passing process. 
    \item Graph Attention Networks (\textbf{GAT}) \cite{Velickovic2018Graph}: a recent model which takes attention mechanism to align different weights to neighbors during the information aggregating process. 
\end{itemize}

\textit{Heterogeneous GNNs (HeteG) Based Methods}
\begin{itemize}
    \item Relational Graph Convolutional Networks (\textbf{RGCN}) \cite{Schlichtkrull2018Modeling}: an advanced extension of GCN, which takes relationship information into consideration by giving different weights for different relationships.
    \item Heterogeneous Graph Neural Network (\textbf{HetGNN})
    \cite{Zhang2019Heterogeneous}: a multi-modal heterogeneous graph model that uses Bi-LSTM to process multi-modal information and then applies the attention mechanism in heterogeneous information fusing.
    \item Heterogeneous Graph Attention Network (\textbf{HAN})
    \cite{Wang2019Heterogeneous}: one of the earliest models to implement hierarchical attention based on the metapath relationships in graph neural networks.
    \item interpretable and efficient Heterogeneous Graph Convolutional Network (\textbf{ie-HGCN})
    \cite{Yang2021Interpretable}: a SOTA model that first implements object-level aggregation and then aggregates type-level information based on different metapaths.
    \item Heterogeneous-attention-network-based model (\textbf{HAT}) \cite{zheng2021heterogeneous}: a SOTA model which conducts triple-level attention in SMEs bankrupt prediction.
\end{itemize}

% \textit{GNNs for Bankruptcy Prediction}

% \begin{itemize}
%      \item Heterogeneous-attention-network-based model (HAT) \cite{zheng2021heterogeneous}: a SOTA model which conducts triple level attention in SMEs bankrupt prediction.
% \end{itemize}
\subsection{Experiment Details}
\label{experiment details}

% For all baseline methods, we calculate enterprise risk information by counting each number of lawsuit attributes and concentrating them with enterprise basic business attributes as enterprise risk representations. We utilize random initialization based on standard normal distribution to assign initial representations for enterprises and persons when implementing GNNs based methods. We choose Metapath2vec \cite{dong2017metapath2vec} as the pre-trained model for \textsc{ComRisk} to generate the supplement embeddings. We implement \textsc{ComRisk} and baselines with PyTorch and PyTorch Geometric (PyG). We refer to THU-HyperG \cite{gao2020hypergraph} for constructing hypergraphs. We implement baselines through official codes with fine-tuning parameters including hidden dimension, layer number and multi-head number to get better performances as much as we can. All neural network based models are trained with Adam optimizer \cite{kingma2014adam} with the Cosine Annealing Learning Rate Scheduler \cite{loshchilov2016sgdr}. We set input dimension 16 and output dimension 12 for each model.
% We run all the methods for 500 epochs and update models considering the improvement of both two comprehensive indicators on validation dataset, i.e., accuracy and F1 score to alleviate overfitting problem. We report the results of all methods on the testing dataset.

For all of the baseline methods, we calculate enterprise risk information by counting the number of lawsuit attributes and combining them with basic enterprise business attributes as enterprise risk representations. We use random initialization based on standard normal distribution to assign initial representations for enterprises and persons when implementing GNN-based methods. We choose Metapath2vec \cite{dong2017metapath2vec} as the pre-trained model for \textsc{ComRisk} to generate the supplement embeddings. We implement \textsc{ComRisk} and baselines with PyTorch and PyTorch Geometric (PyG). We refer to THU-HyperG \cite{gao2020hypergraph} to construct the hypergraphs. We implement baselines based on official codes with fine-tuning parameters, including hidden dimension, layer number, and multihead number, to obtain better performance. All neural network–based models are trained with the Adam optimizer \cite{kingma2014adam} and the Cosine Annealing Learning Rate Scheduler \cite{loshchilov2016sgdr}. We set input dimension 16 and output dimension 12 for each model. We run all methods for 500 epochs and update the models considering the improvement of the two comprehensive indicators on the validation dataset (i.e., the accuracy and F1 score to alleviate the overfitting problem). We report the results of all methods on the testing dataset.

% More details can be found in Appendix \ref{experiment details}.

% For all baseline methods, we calculate enterprise risk information by counting each number of lawsuit attributes and concentrating them with enterprise basic business attributes as enterprise risk representations. Besides, for GNNs based methods, we utilize the same initialization as our model to assign representations for enterprises and persons. We implement \textsc{ComRisk} and baselines with PyTorch and PyTorch Geometric (PyG). We refer to THU-HyperG \cite{gao2020hypergraph} for constructing hypergraphs. We implement baselines through official codes and default parameters. All neural network based models are trained with SGD optimizer with the Cosine Annealing Learning Rate Scheduler \cite{loshchilov2016sgdr}. We set input dimension 64 and output dimension 12 for each model. We run all the methods for 200 epochs and update models considering the improvement of both two comprehensive indicators on validation dataset, i.e., accuracy and F1 score to alleviate overfitting problem. Finally, we report the results of all methods on the testing dataset.

\subsection{Experimental Results and Analysis}

% \textbf{Complexity analysis}

% Table \ref{Table:overall-performance} shows the evaluation results against twelve state-of-the-art baselines, from which we observe that our proposed method outperforms all baselines for enterprise bankruptcy prediction in terms of all comprehensive metrics on our newly generated dataset SMEsD. Specifically, \textsc{ComRisk} achieves state-of-art performance with improvements of 4.68\%, 1.38\% and 9.23\% on accuracy, F1 and AUC score respectively, which confirms the capability of our method in mixing both enterprise intra-risk and contagion risk for bankruptcy prediction.

Table \ref{Table:overall-performance} shows the evaluation results against 12 SOTA baselines. We can see that the proposed method outperforms all baselines for enterprise bankruptcy prediction in terms of all of the comprehensive metrics on our newly generated dataset (SMEsD). Specifically, COMRISK achieves SOTA performance with improvements of 4.68\%, 1.38\%, and 9.23\% for accuracy, F1, and AUC scores, respectively. This confirms the ability of our method to use both intrarisk and contagion risk for bankruptcy prediction.

\begin{table}[htb]
\caption{The overall performance}
\label{Table:overall-performance}
\resizebox{0.49\textwidth}{!}{
% \begin{center} 
\begin{tabular}{ll|c|c|c|c|c}
\toprule
\multicolumn{2}{c|}{\textbf{Models}}                                      & \multicolumn{1}{l|}{\textbf{Accuracy}} & \multicolumn{1}{l|}{\textbf{Precision}} & \multicolumn{1}{l|}{\textbf{Recall}} & \multicolumn{1}{l|}{\textbf{F1}} & \multicolumn{1}{l}{\textbf{AUC}} \\ \hline
\multicolumn{1}{c|}{\multirow{3}{*}{\textbf{ML}}}     & \textbf{LR (2013 \cite{hosmer2013applied})}      & 0.6090                                  & 0.6780                                   & 0.7547                               & 0.7143                           & 0.5812                            \\ 
\multicolumn{1}{c|}{}                                 & \textbf{SVM (1999 \cite{suykens1999least})}     & 0.6314                                 & 0.6612                                  & 0.8836                               & 0.7564                           & 0.5256                            \\ 
\multicolumn{1}{c|}{}                                 & \textbf{GBDT (2001\cite{friedman2001greedy})}    & 0.6456                                 & 0.7449                                  & 0.6887                               & 0.7157                           & 0.6843                            \\ \midrule
\multicolumn{1}{c|}{\multirow{2}{*}{\textbf{HomoG}}}  & \textbf{GCN (2017 \cite{Kipf2017Semi-supervised})}     & 0.6619                                 & 0.6792                                  & 0.9057                               & 0.7763                           & 0.7099                            \\ 
\multicolumn{1}{c|}{}                                 & \textbf{GAT (2018 \cite{Velickovic2018Graph})}     & 0.6802                                 & 0.6998                                  & 0.8868                               & 0.7822                           & 0.6251                            \\ \midrule
\multicolumn{1}{c|}{\multirow{2}{*}{\textbf{HyperG}}} & \textbf{HGNN (2019 \cite{feng2019hypergraph})}    & 0.6884                                 & 0.6941                                  &\underline{\textbf{0.9277} }                              & 0.7941                           & 0.6433                            \\ 
\multicolumn{1}{c|}{}                                 & \textbf{HWNN (2021 \cite{sun2021heterogeneous})}    & 0.6640                                  & 0.7029                                  & 0.8333                               & 0.7626                           & 0.6395                            \\ \midrule
\multicolumn{1}{c|}{\multirow{5}{*}{\textbf{HeteG}}}  & \textbf{RGCN (2018 \cite{Schlichtkrull2018Modeling})}    & 0.6965                                 & 0.7464                                  & 0.8050                                & 0.7746                           & 0.6857                            \\ 
\multicolumn{1}{c|}{}                                 & \textbf{HetGNN (2019 \cite{Zhang2019Heterogeneous})}  & 0.6965                                 & 0.7036                                  & 0.9182                               & 0.7967                           & 0.7185                            \\  
\multicolumn{1}{c|}{}                                 & \textbf{HAN (2019 \cite{Wang2019Heterogeneous})}     & \underline{0.7332}                                  & 0.7429                                  & 0.8994                               & \underline{0.8137}                            & 0.7331                            \\  
\multicolumn{1}{c|}{}                                 & \textbf{ie-HGCN (2021 \cite{Yang2021Interpretable})} & 0.7210                                  & \underline{0.7521}                                   & 0.8491                               & 0.7976                           & \underline{0.7560}
                             \\  
\multicolumn{1}{c|}{}                                 & \textbf{HAT (2021 \cite{zheng2021heterogeneous})}     & \multicolumn{1}{c|}{0.7312}             & \multicolumn{1}{c|}{0.7435}              & \multicolumn{1}{c|}{0.8931}           & \multicolumn{1}{c|}{0.8114}       & \multicolumn{1}{c}{0.7006}        \\ \midrule
\multicolumn{2}{l|}{\textbf{Loss-weighted ComRisk}}                      & \multicolumn{1}{c|}{0.7739}             & \multicolumn{1}{c|}{0.7820}               & \multicolumn{1}{c|}{0.9025}           & \multicolumn{1}{c|}{\textbf{0.8380} }        & \multicolumn{1}{c}{0.8256}        \\ \midrule
\multicolumn{2}{l|}{\textbf{ComRisk}}                                    & \multicolumn{1}{c|}{\textbf{0.7800} }             & \multicolumn{1}{c|}{\textbf{0.8409} }              & \multicolumn{1}{c|}{0.8145}           & \multicolumn{1}{c|}{0.8275}       & \multicolumn{1}{c}{\textbf{0.8483}}        \\ 
\bottomrule
\end{tabular}
}
% \end{center}
% }
\end{table}

{\textbf{Major Analysis.}} 
% (1) We could observe that SVM achieves good performance on recall, that is because lawsuit information as well as enterprise basic information is highly correlated with enterprise bankruptcy. However, SVM achieves poor performance on other comprehensive metrics because of overfitting; (2) We can observe that all the graph based models including hyper-graph neural networks and heterogeneous graph neural networks perform better in most metrics than machine learning methods, which demonstrates the superiority of contagion risk in enterprise bankruptcy prediction; (3) Besieds, we can also observe that HWNN performs better than HGNN because of considering different types of hyperedges  which shows the necessity of considering hypergraph heterogeneity; (4) In addition, we can find that the two SOTA HeterG baseline model, i.e., ie-HGCN and HAN, show better performances than ML and HomoG models, which affirms the ability of heterogeneous graphs in capturing contagion risk.
(1) We observe that the SVM achieves good performance for recall. This is because lawsuit information and basic enterprise information are highly correlated with enterprise bankruptcy. However, the SVM has poor performance on other comprehensive metrics because of overfitting. (2) We can observe that all graph based models, including hypergraph neural networks and heterogeneous graph neural networks, perform better on most metrics than machine learning methods. This demonstrates the superiority of using contagion risk for enterprise bankruptcy prediction. (3) We also find that HWNN performs better than HGNN because it considers different types of hyperedges; this demonstrates the necessity of considering hypergraph heterogeneity. (4) In addition, we find that the two SOTA HeterG baseline models (i.e., ie-HGCN and HAN) show better performance than ML and HomoG models, which confirms the ability of heterogeneous graphs to capture contagion risk.

{\textbf{Credit Scenario Analysis}}.
% Enterprise bankruptcy prediction can be applied in the credit scenario, in which it is of great importance to support SMEs as they contribute a lot to the economy. In previous years, banks prefer not to provide loans for SMEs as they usually don't have reliable access to the SMEs' risk level, resulting in high recall score in loan decision, which is nonsense. On the contrast, precision is a better indicator which can help banks to exclude high-risk enterprises and provide morn loan to SMEs without bring more loss. We can observe from Table \ref{Table:overall-performance} that the proposed model \textsc{ComRisk} achieve 8.88\% gain than SOTA baseline model which can benefit both loan decision makers and SMEs. Besides, to promote the recall score, we propose the \textsc{Loss-weighted ComrRisk} which assigns larger weight for the bankruptcy enterprises' loss during the training process. We can also observe from Table \ref{Table:overall-performance} that the \textsc{Loss-weighted ComrRisk} achieve comparable performance in recall and keep excellent performances in other metrics at same time. 
Enterprise bankruptcy prediction can be applied to the credit scenario. It is important to support SMEs in this regard since they contribute a great deal to the economy. In the past, banks preferred not to give loans to SMEs since the banks usually did not have reliable access to SMEs’ risk levels, resulting in high recall scores in loan decisions. By contrast, precision is a better indicator that can help banks exclude high-risk enterprises and offer more loans to SMEs while avoiding losses. We can see in Table \ref{Table:overall-performance} that the proposed model \textsc{ComRisk} achieves an 8.88\% gain over the SOTA baseline model, which can benefit both loan decision-makers and SMEs. Meanwhile, to promote the recall score, we propose \textsc{Loss-weighted ComrRisk}, which assigns more weight for bankrupt enterprises’ losses during the training process. We can also see in Table \ref{Table:overall-performance} that \textsc{Loss-weighted ComrRisk} achieves comparable performance in recall and maintains excellent performance in other metrics at same time.

\begin{figure}[t]
    \centering
    \includegraphics[width=0.48\textwidth]{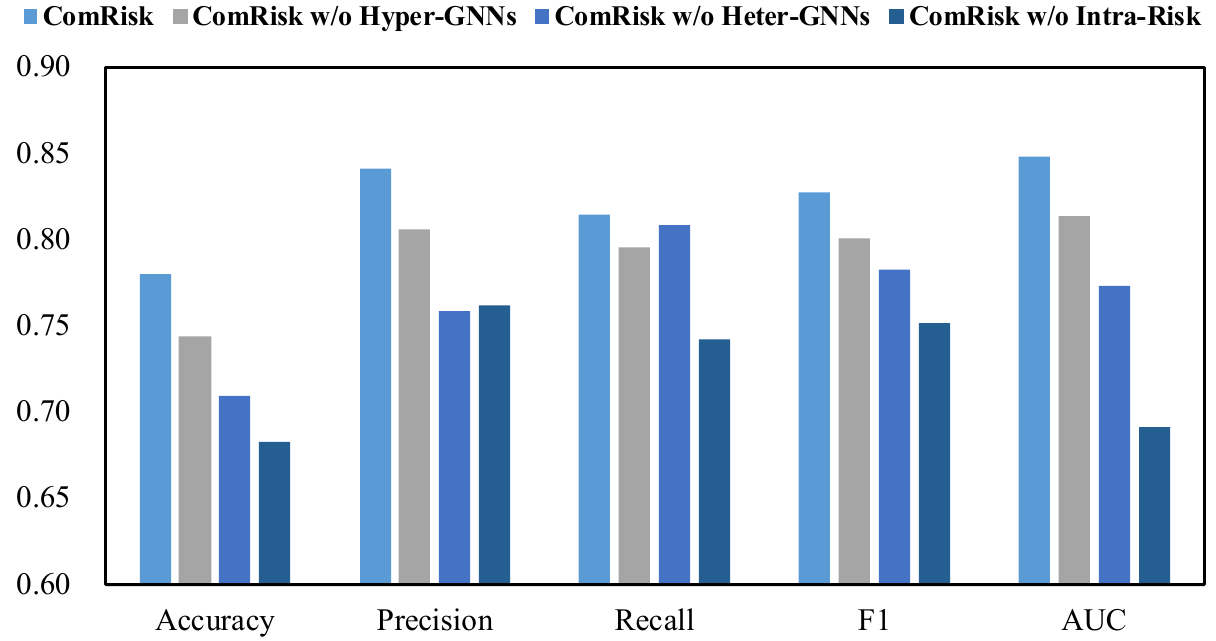}
    \caption{Ablation Study}
    \label{fig:Ablation}
\end{figure}

% and GAT performs better than GCN as a result of attention mechanism.
\subsection{Ablation Study}

We conduct an ablation experiment to evaluate the effectiveness of different components in the proposed model \textsc{ComRisk}. The three ablated variants are as follows: (1) \textbf{\textsc{ComRisk} w/o Intra-Risk}, which deletes the inner risk encoder; (2) \textbf{\textsc{ComRisk} w/o Hyper-GNNs}, which removes the hierarchical hypergraph encoder; and (3) \textbf{\textsc{ComRisk} w/o Heter-GNNs}, which deletes the hierarchical risk encoder module. Figure \ref{fig:Ablation} shows the results. We can see that removing either the heterogeneous graph, hypergraph, or risk encoder leads to performance degeneration, which demonstrates the effectiveness of the three modules. Specifically, the proposed model \textsc{ComRisk} outperforms \textbf{\textsc{ComRisk} w/o Intra-Risk}, which confirms the effectiveness of lawsuit information for bankruptcy prediction. Meanwhile, \textbf{\textsc{ComRisk} w/o Intra-Risk} has the worst performance among the three ablated variants, which verifies the importance of intrarisk information. Thus, we highlight the design of capturing lawsuit risk information. Compared with \textbf{\textsc{ComRisk} w/o Hyper-GNNs}, the proposed model \textsc{ComRisk} also achieves better performance, which demonstrates the contribution of hypergraphs. This is because enterprises in the same industry, in the same area, or with same stakeholders usually face similar external risks (e.g., industry development recession, regional economic policy changes, and guarantee risks), which can be detected by hypergraphs. For \textbf{\textsc{ComRisk} w/o Heter-GNNs}, we find that performance also decreases, which confirms that utilizing complex heterogeneous relationships in an EKG can strengthen the capacity of the model.

\begin{figure}[t]
    \centering
    \includegraphics[width=0.48\textwidth]{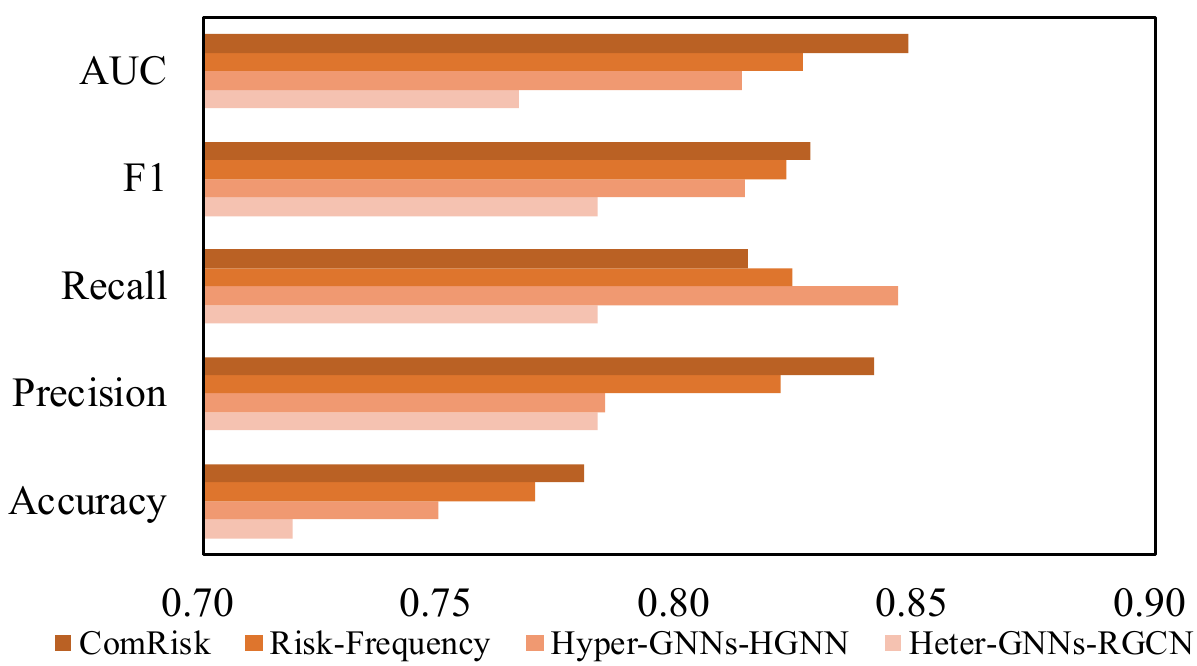}
    \caption{Variants Analysis}
    \label{fig:Variants}
\end{figure}

\subsection{Variant Analysis }

We conduct a variant analysis of \textsc{ComRisk} to show the effectiveness of its architecture. (1) \textbf{\textsc{ComRisk}-Frequency} replaces the proposed inner risk encoder with the frequency of lawsuit attributes with regard to each enterprise; (2) \textbf{Hyper-GNNs-HGNN} replaces the hierarchical hypergraph encoder with HGNN; and (3) \textbf{Heter-GNNs-RGCN} uses RGCN rather than the proposed hierarchical risk encoder. Figure \ref{fig:Variants} shows the results. We can see that the proposed \textsc{ComRisk} achieves the best performance compared to all variants. Specifically, \textsc{ComRisk} performs better than Risk-Frequency, which again demonstrates the risk-representation capacity of the inner risk encoder. This is because our model not only uses lawsuit risk information in terms of frequency but also considers the time interval related to each lawsuit, which is shown to be significantly correlated with enterprise bankruptcy in Table \ref{tab:bankruptcy-analysis}. Compared with Hyper-HGNN, the proposed model \textsc{ComRisk} also performs better because it can distinguish different types of hyperedges and assign different importance weights for the learned representations. We also observe that replacing the hierarchical risk encoder with RGCN lowers performance, from which we can conclude that the proposed hierarchical risk encoder can better capture the contagion risk embedded in complex relationships.

\begin{figure}[t]
    \centering
    \includegraphics[width=0.3\textwidth]{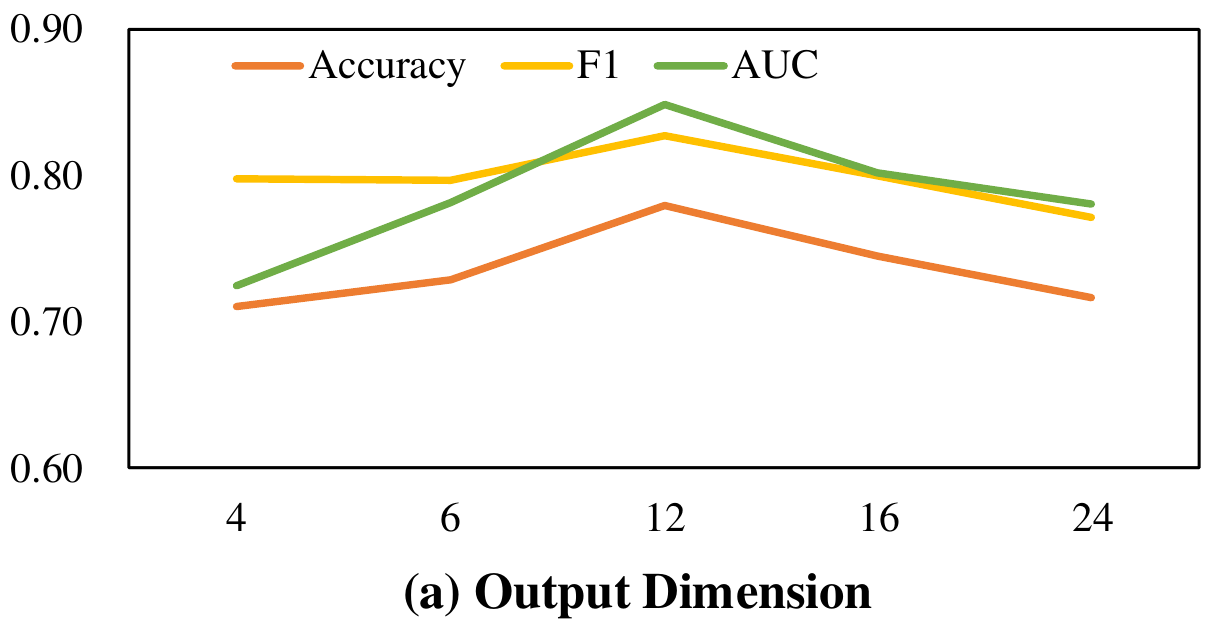}
    \includegraphics[width=0.3\textwidth]{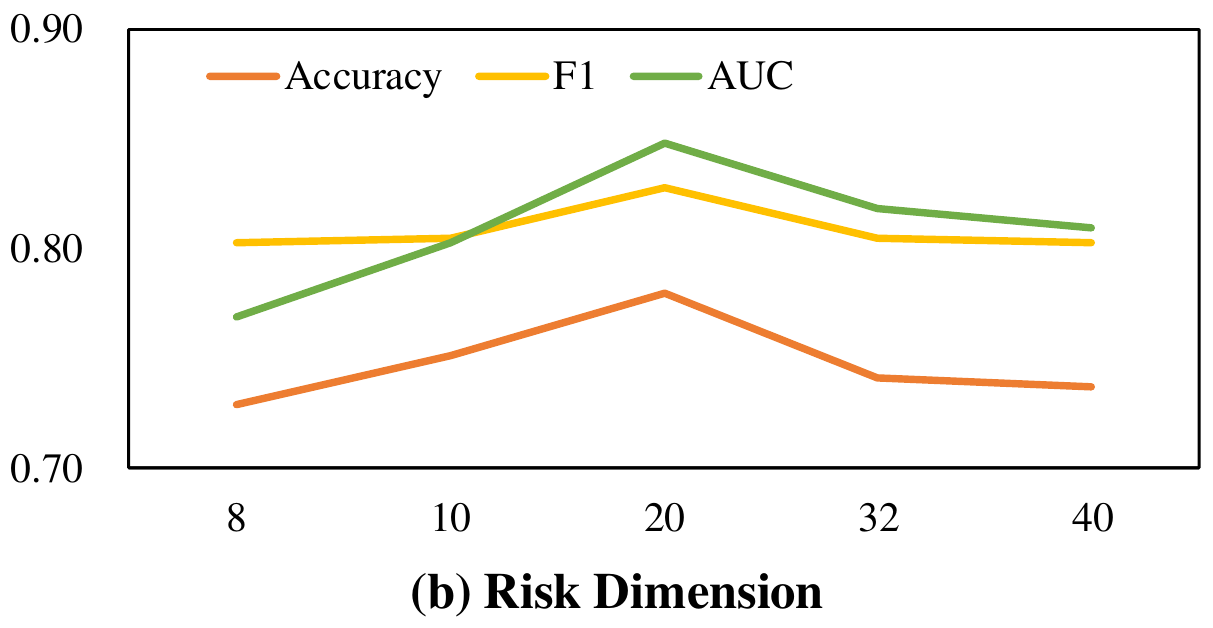}
    \caption{Parameter Analysis}
    \label{fig:parameter-analysis}
\end{figure}

\subsection{Parameter Analysis}
% We examine the effects of the two critical hyper-parameters, i.e., output dimension and lawsuit risk information dimension of \textsc{ComRisk}, the default dimension of which are 12 and 20 respectively. 
% We measure the performance with two comprehensive indicators accuracy and F1 score.

We examine the effects of the two critical hyper-parameters (i.e., the output dimension and lawsuit risk information dimension in \textsc{ComRisk}), the default dimensions of which are 12 and 20, respectively.

\textbf{Impact of input dimension.} 
% As shown in Figure \ref{fig:parameter-analysis} (a), we can find that the performance firstly rises up with the dimension increasing before 12 and then falls with the dimension increasing. This is may because a model with too low dimension fails to represent abundant information of nodes. Meanwhile, high dimension brings too much noisy information and thus restricts the capacity of the proposed model \textsc{ComRisk}.
As shown in Figure \ref{fig:parameter-analysis} (a), performance first increases with the dimension increasing before 12 and then falls with the dimension increasing. This could be because a model with a too-low dimension fails to represent abundant node information. Meanwhile, a high dimension produces too much noisy information and thus restricts the capacity of the model (\textsc{ComRisk}).

\textbf{Impact of lawsuit risk information dimension.} 
% We can observe from Figure \ref{fig:parameter-analysis} (b) that the model performance firstly increases and reaches its peak in 20, then decreases with the dimension rising up. This is mainly because the number of total lawsuit attributes in the SMEsD dataset is 20, both lower and higher lawsuit risk dimension leads to performance decrease. 
We can see in Figure \ref{fig:parameter-analysis} (b) that model performance first increases and reaches its peak at 20 and then decreases with the dimension rising. This is mainly because the number of total lawsuit attributes in the SMEsD dataset is 20; lower and higher lawsuit risk dimensions both lead to a decrease in performance.

\section{Conclusion}
\label{section}
% In this paper, we propose to model enterprise bankruptcy risk by combining its intra-risk with contagion risk. Under this framework, we propose a novel method that is equipped with an intra-risk encoder and GNNs-based contagion risk encoder. Specifically, the intra-risk encoder is able to capture enterprise intra-risk using the statistic correlated indicators from the basic business information and litigation information. The contagion risk encoder consists of hypergraph neural networks and heterogeneous graph neural networks, which aim to model contagion risk through two aspects, i.e. hyperedge and multiplex heterogeneous relations among enterprise knowledge graph, respectively. 
% To evaluate the proposed model, we collect multi-sources SMEs data and build a new dataset SMEsD, on which the experimental results demonstrate the superiority of the proposed method. The dataset is expected to become a significant benchmark dataset for SMEs bankruptcy prediction and promote the development of financial risk study further.

In this study, we propose modeling enterprise bankruptcy risk by combining intrarisk and contagion risk. In this framework, we propose a novel method that includes an intrarisk encoder and GNNs based contagion risk encoder. Specifically, the intrarisk encoder can capture enterprise intrarisk using statistically correlated indicators derived from basic business information and litigation information. The contagion risk encoder consists of hypergraph neural networks and heterogeneous graph neural networks, which aim to model contagion risk in the two aspects of hyperedge and complex heterogeneous relationships among EKGs, respectively. To evaluate the proposed model, we collect multisource SME data and build a new dataset, SMEsD. The experimental results demonstrate the superiority of the proposed method. The dataset is expected to become a significant benchmark dataset for SME bankruptcy prediction while further promoting research on financial risk.

% \appendices

% \section{}

\ifCLASSOPTIONcompsoc
  \section*{Acknowledgments}
\else
  \section*{Acknowledgment}
\fi

The authors would like to thank all anonymous reviewers in advance.
This research has been partially supported by grants from the National Natural Science Foundation of China under Grant No. 71725001, 71910107002, 61906159, 62176014, U1836206, 71671141, 71873108, 62072379, the State key R \& D Program of China under Grant No. 2020YFC0832702, the major project of the National Social Science Foundation of China under Grant No. 19ZDA092, the Financial Intelligence and Financial Engineering Key Laboratory of Sichuan Province and the  Fundamental Research Funds for the Central Universities under Grant No. JBK2207004.

\ifCLASSOPTIONcaptionsoff
  \newpage
\fi



\bibliographystyle{IEEEtran}


% H.~Kopka and P.~W. Daly, \emph{A Guide to {\LaTeX}}, 3rd~ed.\hskip 1em plus

% \end{thebibliography}

\bibliography{sample-base}


% \begin{IEEEbiography}[{\includegraphics[width=1in,height=1.25in,clip,keepaspectratio]{mshell}}]{Michael Shell}

\begin{IEEEbiography}[{\includegraphics[width=1in,height=1.25in,clip,keepaspectratio]{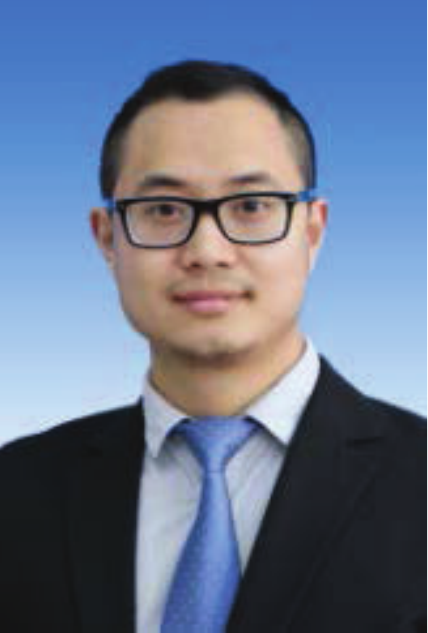}}]{Yu Zhao}
received the B.S. degree from Southwest Jiaotong University in 2006, and the M.S. and Ph.D. degrees from the Beijing University of Posts and Telecommunications in 2011 and 2017, respectively. He is currently an Associate Professor at Southwestern University of Finance and Economics. His current research interests include machine learning, natural language processing, knowledge graph, Fintech. He has authored more than 30 papers in top journals and conferences including IEEE TKDE, IEEE TNNLS, IEEE TMC, ACL.
\end{IEEEbiography}

\begin{IEEEbiography}[{\includegraphics[width=1in,height=1.25in,clip,keepaspectratio]{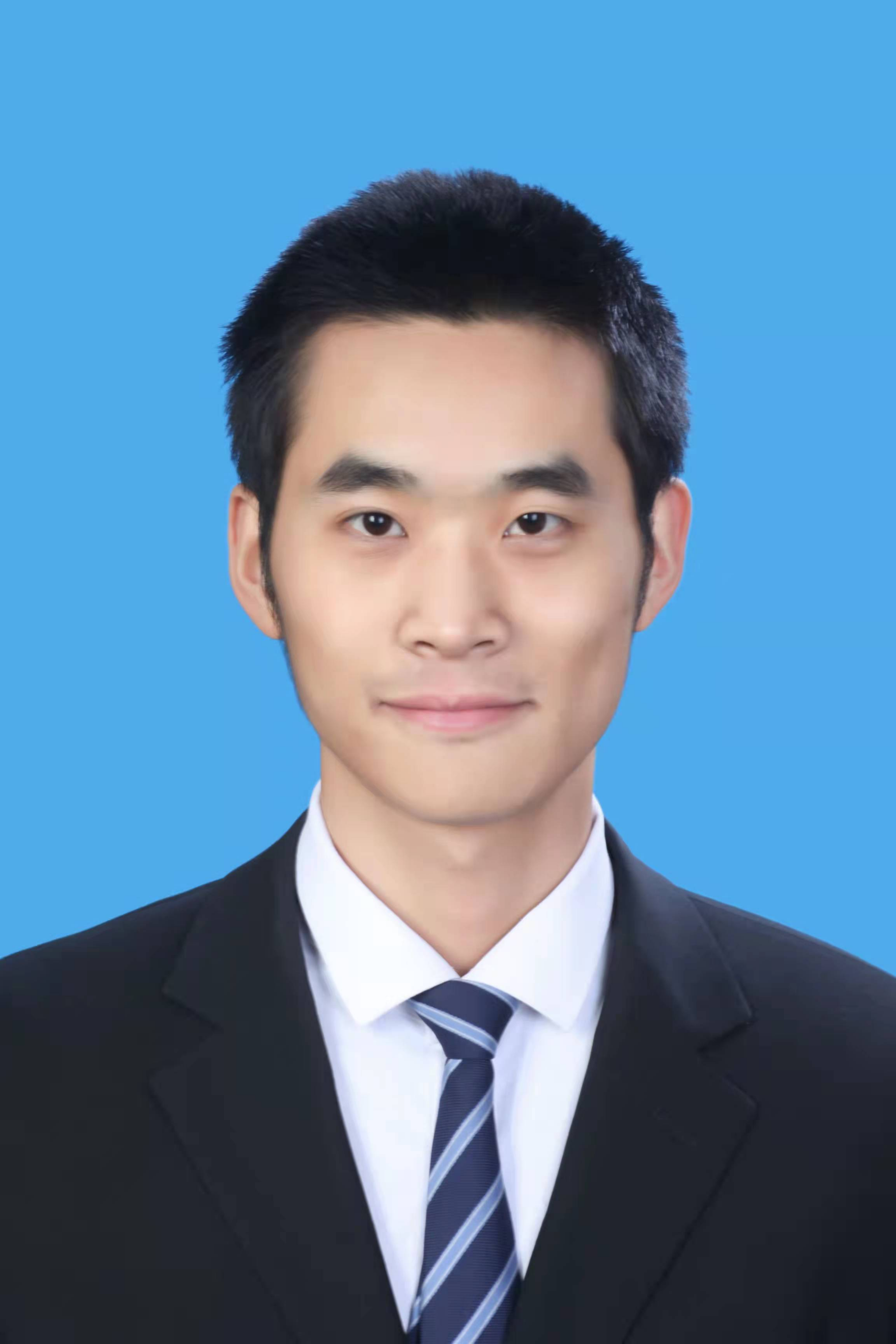}}]{Shaopeng Wei} received the B.S. degree from Huazhong Agricultural University in 2019, and now is a Ph.D student in Southwestern University of Finance and Economics. His research interests include graph learning and relevant applications in Fintech and recommendation system.
% received the B.S. degree from Southwestern University of Finance and Economics in 2018, the M.S. degree in computer science from the University of Hong Kong. She is a data mining engineer in BAIDU. Her research interests include data mining and natural language processing.
\end{IEEEbiography}

\begin{IEEEbiography}[{\includegraphics[width=1in,height=1.25in,clip,keepaspectratio]{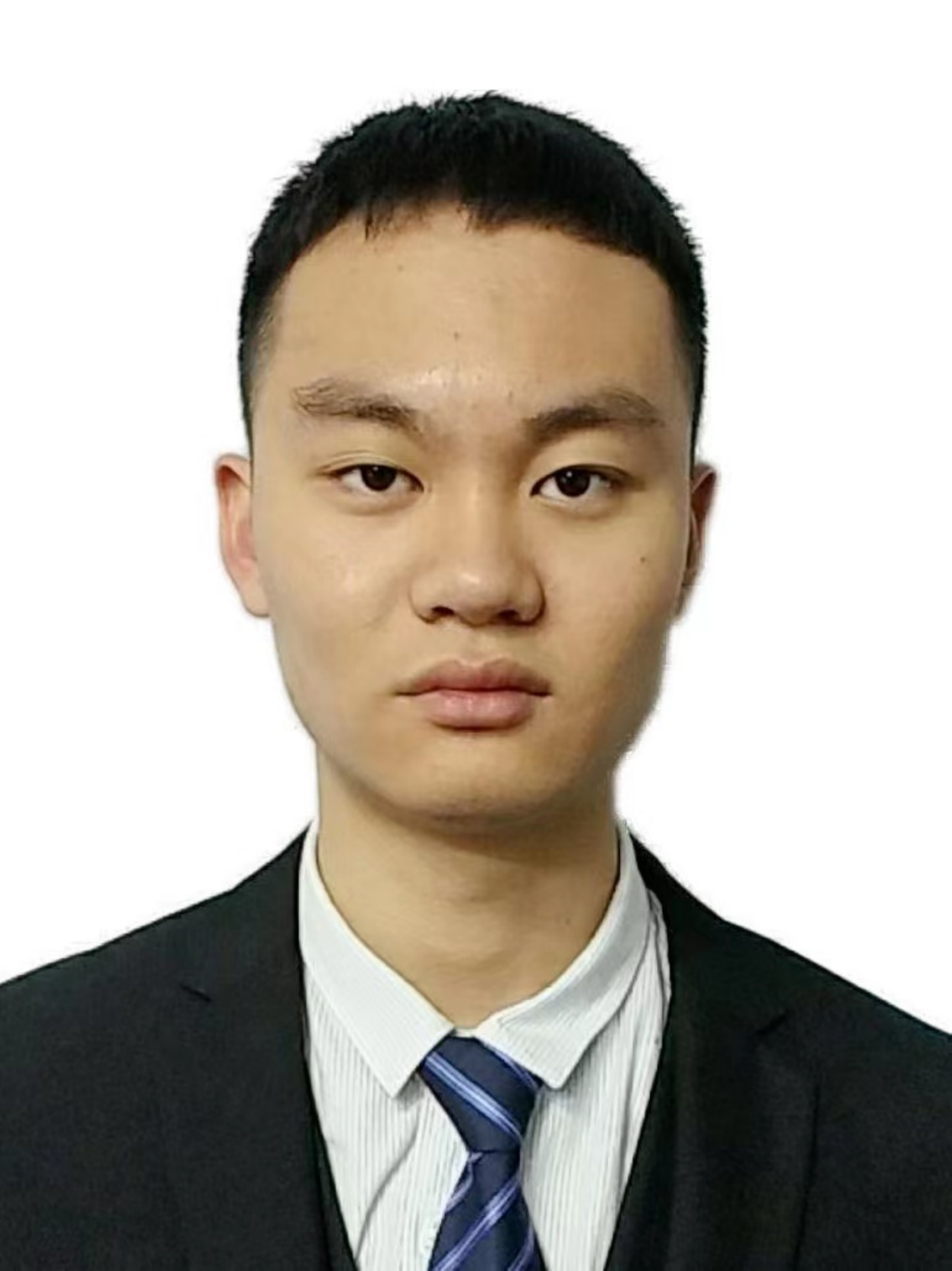}}]{Yu Guo} 
received the B.S. degree from Chengdu Normal University in 2020, and now is a master candidate in Southwestern University of Finance and Economics.  His research interests include natural language processing, dialogue systems, and deep learning.
\end{IEEEbiography}

\begin{IEEEbiography}[{\includegraphics[width=1in,height=1.25in,clip,keepaspectratio]{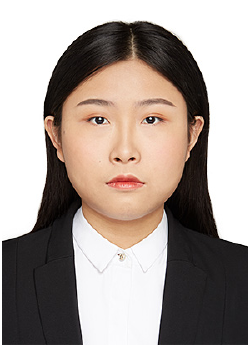}}]{Qing Yang} 
received the B.S. degree from Southwestern University of Finance and Economics in 2021, and now is a master candidate in Southwestern University of Finance and Economics. Her research interests include enterprise risk forecasting.
\end{IEEEbiography}

\begin{IEEEbiography}[{\includegraphics[width=1in,height=1.25in,clip,keepaspectratio]{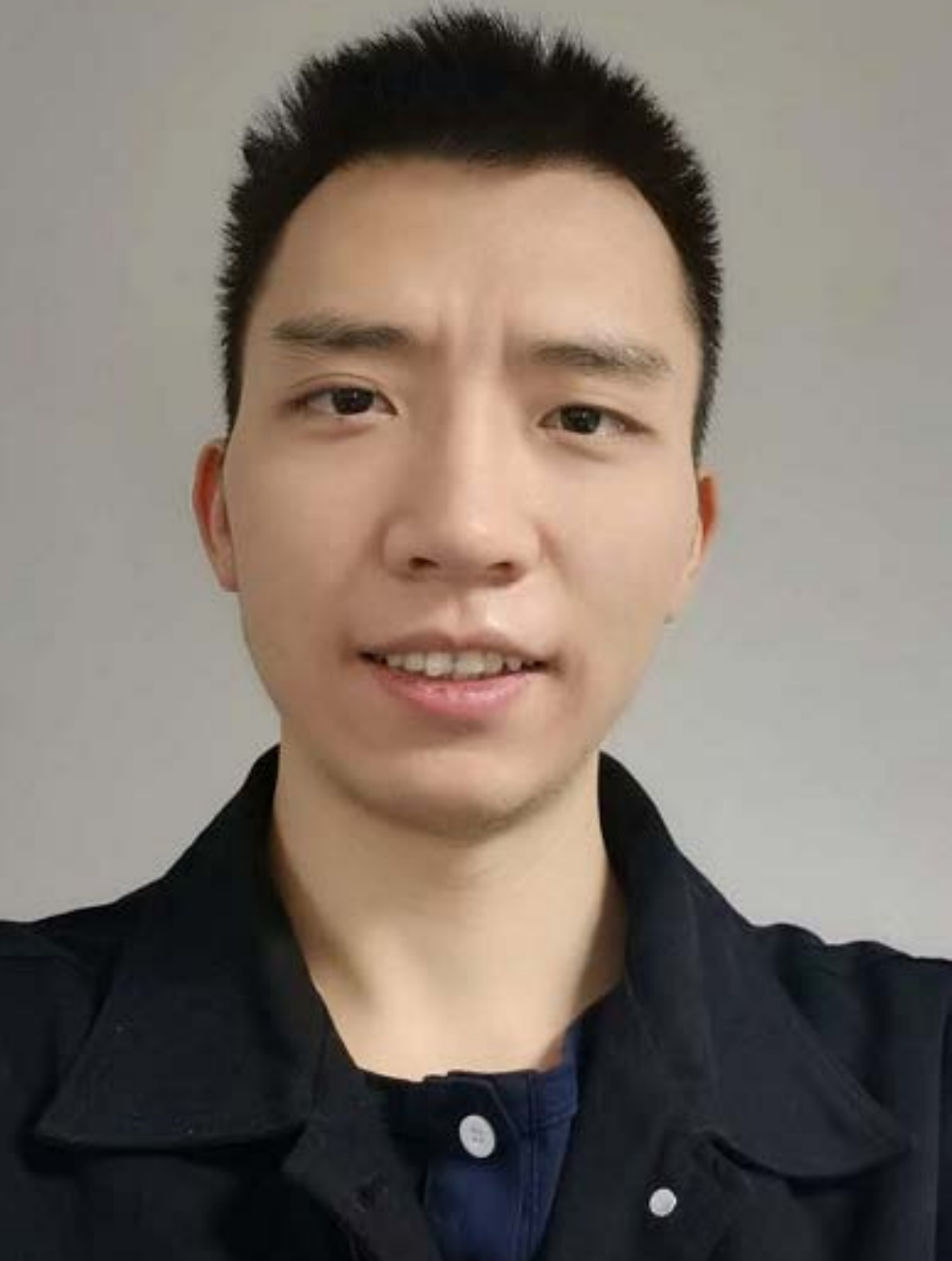}}]{Xingyan Chen}
	received the Ph. D degree in computer technology from Beijing University of Posts and Telecommunications (BUPT), in 2021. 
	He is currently a lecturer with the School of Economic Information Engineering, Southwestern University of Finance and Economics, Chengdu.
	He has published papers in well-archived international journals and proceedings, such as the \textsc{IEEE Transactions on Mobile Computing}, \textsc{IEEE Transactions on Circuits and Systems for Video Technology}, \textsc{IEEE Transactions on Industrial Informatics}, and \textsc{IEEE INFOCOM} etc. 
	His research interests include Multimedia Communications, Multi-agent Reinforcement Learning and Stochastic Optimization.
\end{IEEEbiography}

\begin{IEEEbiography}[{\includegraphics[width=1in,height=1.25in,clip,keepaspectratio]{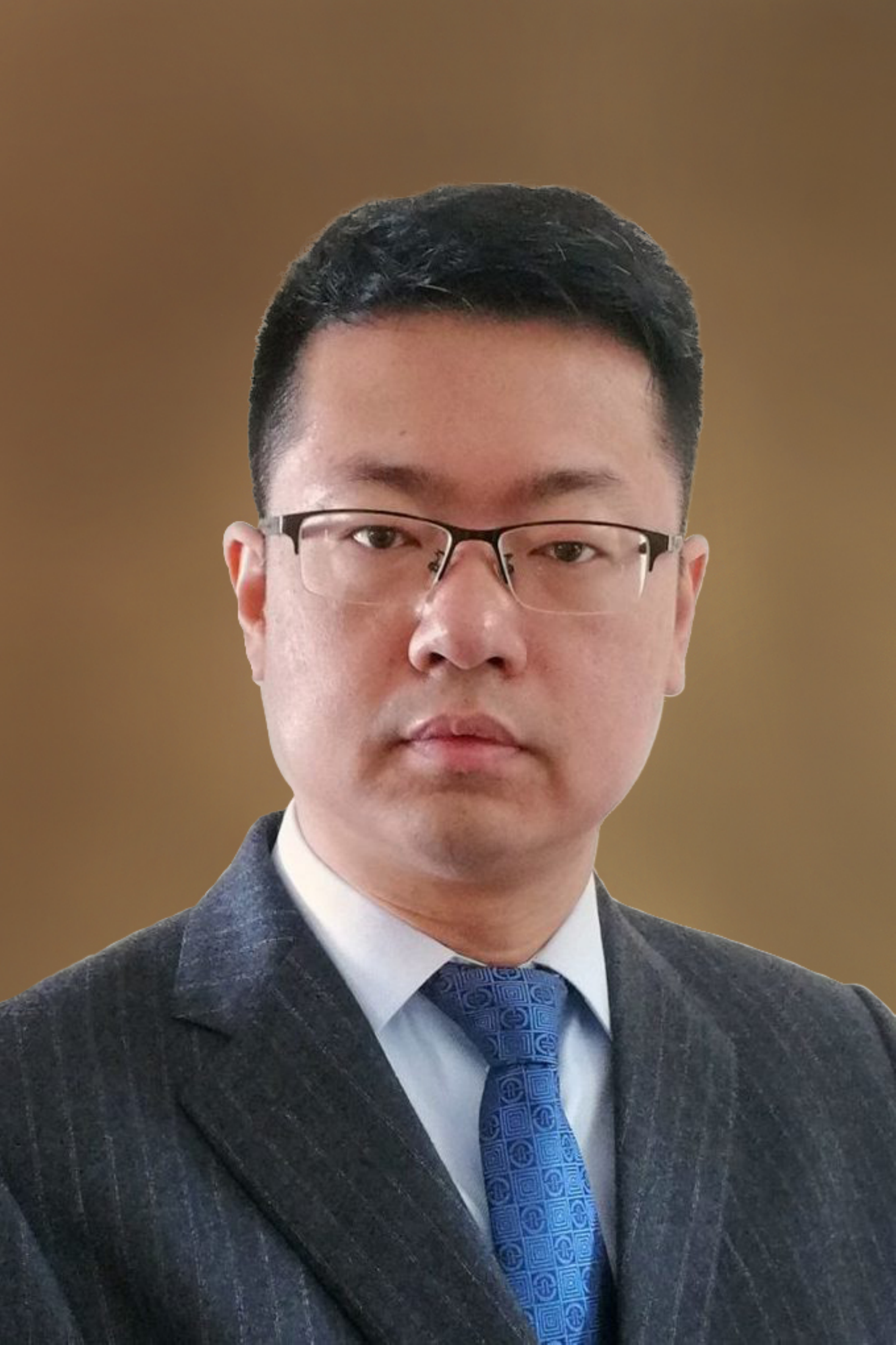}}]{Qing Li} received his PhD degree from Kumoh National Institute of Technology in February of 2005, Korea, and his M.S. and B.S. degrees from Harbin Engineering University, China. He is a postdoctoral researcher at Arizona State University and the Information \& Communications University of Korea. He is a professor at Southwestern University of Finance and Economics, China. His research interests include natural language processing, FinTech. He has published more than 70 papers in the prestigious refereed conferences and journals, such as IEEE TKDE, ACM TOIS, AAAI, SIGIR, ACL, WWW, etc.
\end{IEEEbiography}

\begin{IEEEbiography}[{\includegraphics[width=1in,height=1.25in,clip,keepaspectratio]{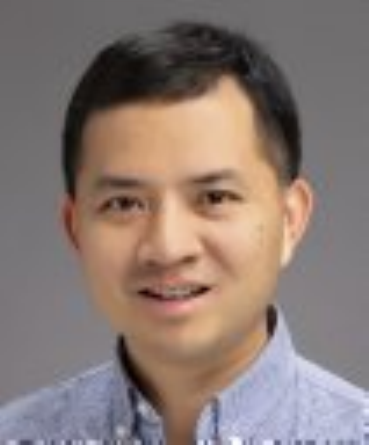}}]{Fuzhen Zhuang} received the PhD degree in computer science from the Institute of Computing Technology, Chinese Academy of Sciences. He is currently a full Professor in Institute of Artificial Intelligence, Beihang University., Beijing 100191, China. His research interests include Machine Learning and Data Mining, including Transfer Learning, Multi-task Learning, Multi-view Learning and Recommendation Systems. He has published more than 100 papers in the prestigious refereed conferences and journals, such as KDD, WWW, SIGIR, ICDE, IJCAI, AAAI, EMNLP, Nature Communications, IEEE TKDE, ACM TKDD, IEEE T-CYB, IEEE TNNLS, ACM TIST, etc.
% received the PhD degree in computer science from the Institute of Computing Technology, Chinese Academy of Sciences. He is an associate professor in the Institute of Artificial Intelligence, Beihang University, Beijing 100191, China. His research interests include transfer learning, machine learning, data mining, distributed classification and clustering, and natural language processing.
\end{IEEEbiography}

\begin{IEEEbiography}[{\includegraphics[width=1in,height=1.25in,clip,keepaspectratio]{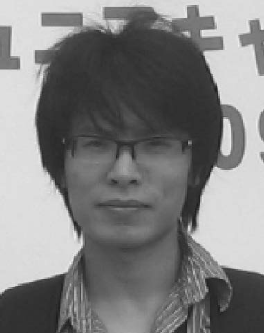}}]{Ji Liu}
received the B.S. degree from the University of Science and Technology of China, Hefei, China, in 2005, the master’s degree from Arizona State University, Tempe, AZ, USA, in 2010, and the Ph.D. degree from the University of Wisconsin–Madison, Madison, WI, USA, in 2014. He is currently an Assistant Professor of computer science, electrical and computer engineering with the Goergen Institute for Data Science, University of Rochester (UR), Rochester, NY, USA, where he created the Machine Learning and Optimization Group. 
% His current research interests include machine learning, optimization, and their applications in other areas such as data mining, healthcare, bioinformatics, computer vision, and many other data analysis involved areas, asynchronous parallel optimization, sparse learning (compressed sensing) theory and algorithm, structural model estimation, online learning, abnormal event detection, and feature/pattern extraction.
He has authored more than 70 papers in top journals and conferences including JMLR, TPAMI, TNNLS, TKDD, NIPS, ICML, SIGKDD, ICCV, and CVPR. Dr. Liu was a recipient of the Award of Best Paper Honorable Mention at SIGKDD 2010, the Award of Best Student Paper Award at UAI 2015, and the IBM Faculty Award. He is named one MIT technology review’s “35 innovators under 35 in China.”
\end{IEEEbiography}

% \begin{IEEEbiography}[{\includegraphics[width=1in,height=1.25in,clip,keepaspectratio]{figure/cezhang.pdf}}]{Ce Zhang} is an assistant professor in Computer Science at ETH Zurich. Before joining ETH, Ce finished his PhD at the University of Wisconsin-Madison and spent another year as a postdoctoral researcher at Stanford, both advised by Christopher Ré. His work has received recongitions such as the SIGMOD Best Paper Award, SIGMOD Research Highlight Award, Google Focused Research Award, an ERC Starting Grant, and has been featured and reported by Science, Nature, the Communications of the ACM, and a various media outlets such as Atlantic, WIRED, Quanta Magazine, etc.
% \end{IEEEbiography}

\begin{IEEEbiography}[{\includegraphics[width=1in,height=1.25in,clip,keepaspectratio]{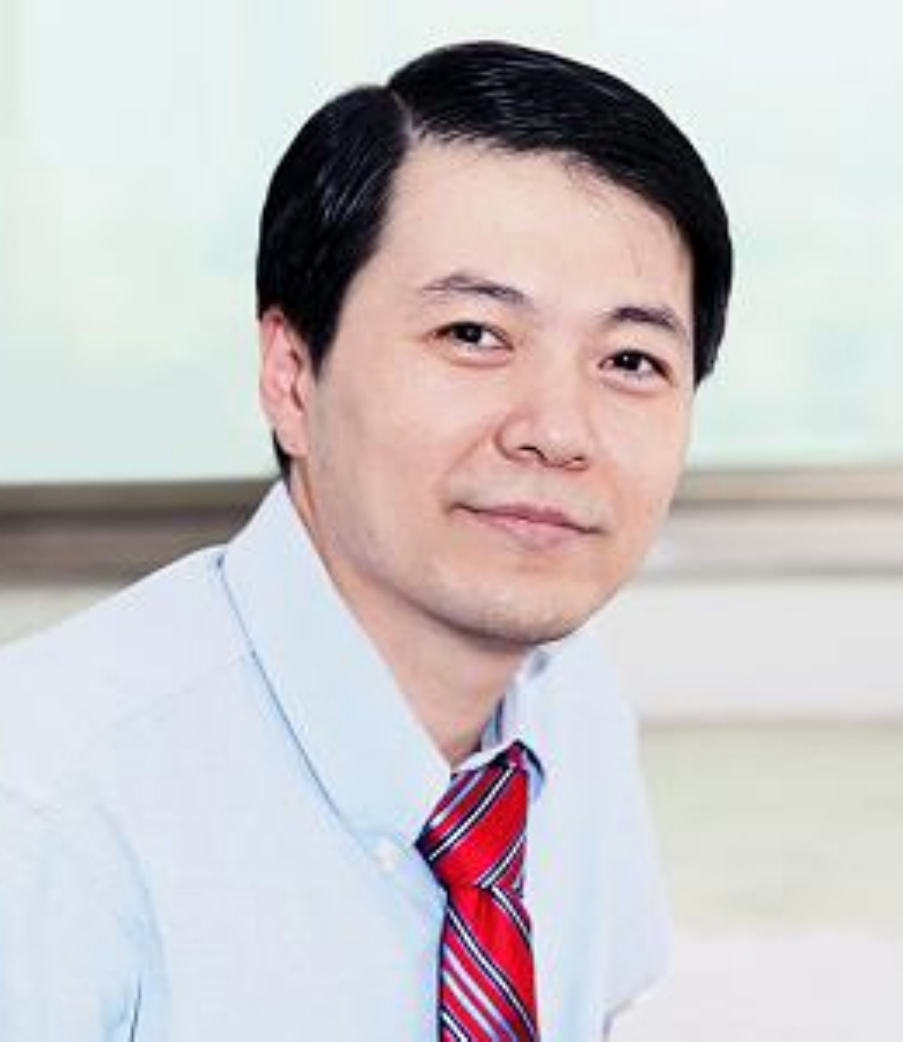}}]{Gang Kou} is a Distinguished Professor of Chang Jiang Scholars Program in Southwestern University of Finance and Economics, managing editor of International Journal of Information Technology \& Decision Making (SCI) and managing editor-in-chief of Financial Innovation (SSCI). He is also editors for other journals, such as: Decision Support Systems, and European Journal of Operational Research. Previously, he was a professor of School of Management and Economics, University of Electronic Science and Technology of China, and a research scientist in Thomson Co., R \& D. He received his Ph.D. in Information Technology from the College of Information Science \& Technology, Univ. of Nebraska at Omaha; Master degree in Dept of Computer Science, Univ. of Nebraska at Omaha; and B.S. degree in Department of Physics, Tsinghua University, China. He has published more than 100 papers in various peer-reviewed journals. Gang Kou’s h-index is 57 and his papers have been cited for more than 10000 times. He is listed as the Highly Cited Researcher by Clarivate Analytics (Web of Science).
% received the PhD degree in computer science from the Institute of Computing Technology, Chinese Academy of Sciences. He is an associate professor in the Institute of Artificial Intelligence, Beihang University, Beijing 100191, China. His research interests include transfer learning, machine learning, data mining, distributed classification and clustering, and natural language processing.
\end{IEEEbiography}

% \begin{IEEEbiography}[{\includegraphics[width=1in,height=1.25in,clip,keepaspectratio]{junguo.png}}]{Jun Guo}
% received the BE and ME degrees from the Beijing University of Posts and Telecommunications (BUPT), China, in 1982 and 1985, respectively, and the PhD degree from the Tohuku Gakuin University, Japan, in 1993. He is currently a professor and a vice president with BUPT. He has authored more than 200 papers in journals and conferences, including Science, Nature Scientific Reports, the IEEE Transactions on PAMI, Pattern Recognition, AAAI, CVPR, ICCV, and SIGIR. His research interests include pattern recognition theory and application, information retrieval, content-based information security, and bioinformatics.
% \end{IEEEbiography}

% \end{IEEEbiography}


%\newpage



%\vfill


\end{document}